\documentclass[manuscript,10pt]{aastex}

\shorttitle{Radio Jets in SS\,433}
\shortauthors{Roberts et al.} 

\begin{document}

\title{Structure and Magnetic Fields in the Precessing \\ Jet System SS\,433  II. Intrinsic Brightness of the Jets}

\author{David H. Roberts, John F. C. Wardle, Michael R. Bell, \\ Matthew R. Mallory, Valerie V. Marchenko, \& Phoebe U. Sanderbeck}
\affil{Department of Physics MS-057, Brandeis University, Waltham, MA 02454-0911 USA}
\email{roberts@brandeis.edu, wardle@brandeis.edu}

\begin{abstract}

Deep Very Large Array imaging of the binary X-ray source SS\,433, sometimes classified as a microquasar, has been used to study the intrinsic brightness distribution and evolution of its radio jets. The intrinsic brightness of the jets as a function of age at emission of the jet material $\tau$ is recovered by removal of the Doppler boosting and projection effects. We find that intrinsically the two jets are remarkably similar when compared for equal $\tau$, and that they are best described by Doppler boosting of the form $D^{2+\alpha}$, as expected for continuous jets. The intrinsic brightnesses of the jets as functions of age behave in complex ways. In the age range $60 < \tau < 150$~days, the jet decays are best represented by exponential functions of $\tau$, but linear or power law functions are not statistically excluded. This is followed by a region out to $\tau \simeq 250$~days during which the intrinsic brightness is essentially constant. At later times the jet decay can be fit roughly as exponential or power law functions of $\tau$.

\end{abstract}


\section{INTRODUCTION}

The galactic X-ray binary source SS\,433, consisting of a stellar-mass black hole in close orbit about an early-type star \citep{BBS2008,HG2008,B2010}, is a miniature analogue of an AGN \citep{MR99}, and is often classified as a microquasar. Two mildly relativistic jets emerge from opposite sides of the compact object at speeds of $0.26~c$. Modeling of the optical spectrum shows that the jet system precesses with a period of 162 days about a cone of half-angle $20^{\circ}$ \citep{AM79, Fabian, Milgrom, M84}. Imaging by the Very Large Array (VLA) confirmed this picture, as the radio images showed helical jets on both sides of the source \citep{Spencer79,HJ81a,HJ81b}. Higher resolution images made by VLBI reveal the structure down to a scale of a few AU \citep{Ver87,Ver93}. Analysis of the 15~GHz VLA-scale structure of the jets in SS\,433, with angular resolution of about $0.1\arcsec$,  was presented in \citet{PaperI}. Multi-epoch dual-frequency analysis of SS\,433 during the summer of 2003 will be presented in \citet{PaperIII}, hereafter Paper~III, and for the summer of 2007 in \citet{PaperIV}, hereafter Paper~IV.

In this paper we use high dynamic range VLA images of SS\,433 to study the radiative intensity of the two jets as a function of the material's {\em birth epoch} $t$ and {\em age at emission}  $\tau$ (hereafter simply ``age''); see Appendix~1 (\S\ref{s:app1}) for our definitions of these quantities.  
Our goals are (i) to determine if the two jets are intrinsically the same, and (ii) to learn if the jets behave as individual non-interacting components or as a continuous stream. SS\,433 offers a unique opportunity to answer these questions because it presents two jets with ever-changing mildly-relativistic velocities known as functions of time and position on the sky from their optical properties. In \S\ref{s:obs} we describe the observations and data reduction. In \S\ref{s:jets} we determine the properties of the jets and we discuss the physical implications of these results in \S\ref{s:Discussion}. Our conclusions are summarized in \S\ref{s:conclude}.

\section{OBSERVATIONS AND DATA REDUCTION}
\label{s:obs}

Interferometer data for 2003 July 11 (JD 2452832) were obtained from the VLA data archive. The array was in the A configuration with 27 working antennas. The frequency used was 4.86~GHz, with a bandwidth of 50~MHz for each of the two IF systems. The data were edited and phase calibrated in AIPS \citep{aips} using the nearby source 1922+155, and amplitude calibrated against 3C\,48; the data were then imaged using the tasks IMAGR and CALIB, utilizing many cycles of phase and amplitude self-calibration. 
These data were previously used by \cite{BB04} to study the symmetries of the jets, and by \cite{MJ08} to study the magnetic field configuration in the jets. Figs.~\ref{fig:CUN}(a) \& (b) show the distribution of total intensity for SS\,433, made with uniform weighting (ROBUST = $-5$); this image has resolution of $0.32\arcsec$. The kinematic model without nutation or velocity variations is shown in (a),  and the model with nutation and velocity variations is shown in (b).  The kinematic model used to define the jet locus is described in  Appendix~1, \S\ref{s:app1}. A naturally weighted image with a resolution of $ 0.47\arcsec$ (ROBUST = +5), shown in Figure~\ref{fig:CNA}, reveals very low level emission out to distances of at least $6\arcsec$, corresponding to ages of about 800~days for both jets.

For all of the analysis that follows we used the image shown Fig.\ref{fig:CUN}, and included the velocity variations and nutation. This accounts for the ``spiking'' apparent in many of the figures. Uncertainty in the position of the model jet locus is by far the largest source of fluctuations in the total intensity curves; the root-mean-square difference between profiles generated with and without these terms is about 1.5~mJy/beam. Thermal noise is negligible until $I_\nu$ falls below about 0.1~mJy/beam. 

Comparison of the images with the kinematic model shows them to be compatible, with the kinematic locus being the leading edge of the jets; the source also contains significant off-jet material. We see no unambiguous evidence of either nutation or significant jet velocity variations, but this is not surprising given the limited resolution of the images.

\section{PROPERTIES OF THE SS\,433 JETS}
\label{s:jets}

\subsection{Normalization Factors}
\label{s:NormalizationFactors}

\begin{figure}[t]
\plotone{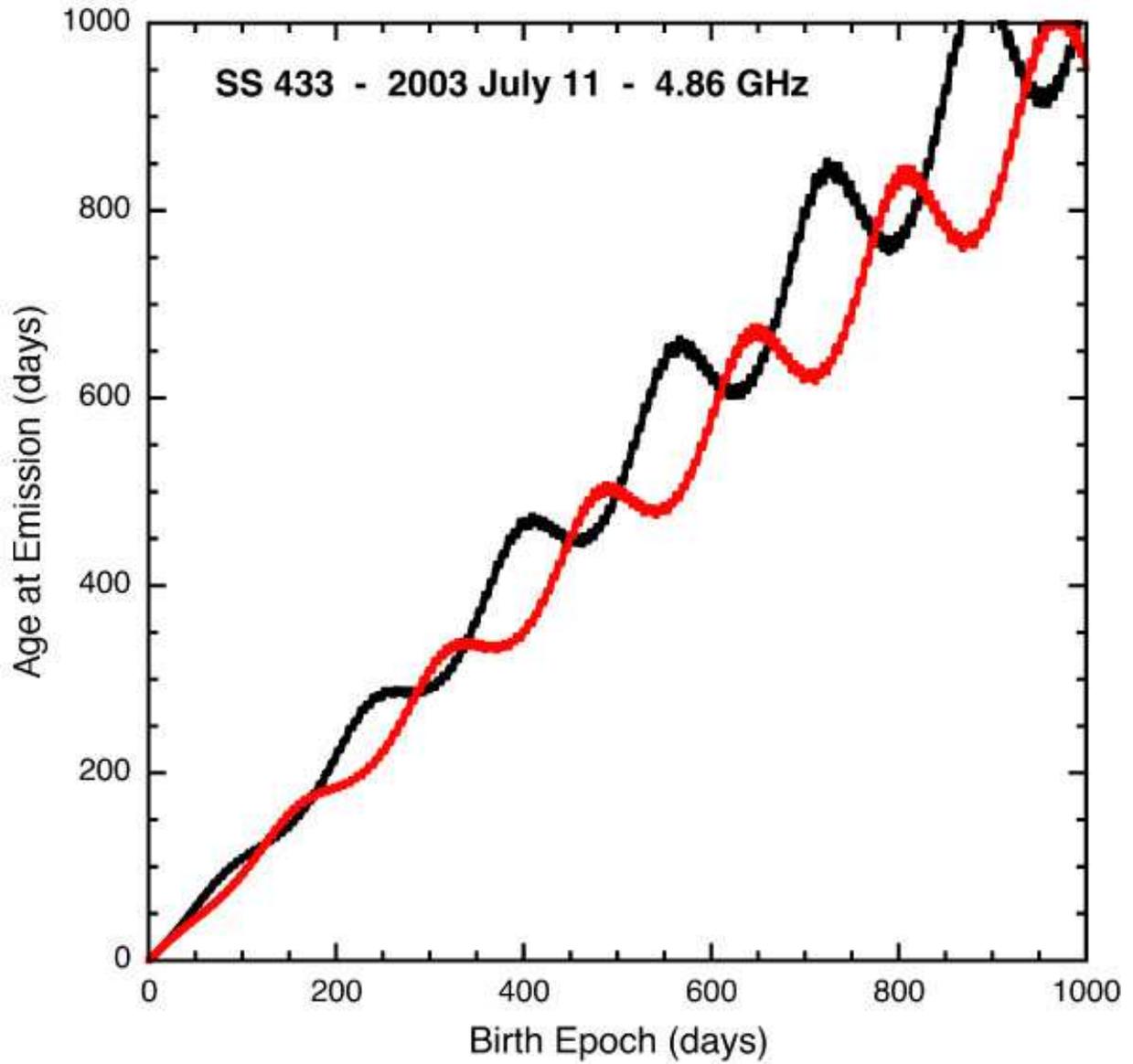}
\caption[]{Light travel time effects in SS\,433. The curves show the age of material in the jets as a function of the birth epoch of the material. The black line is for the east jet, the red line (grey in black \& white) for the west jet.
\label{fig:LTT}}
\end{figure}

\begin{figure}
\plotone{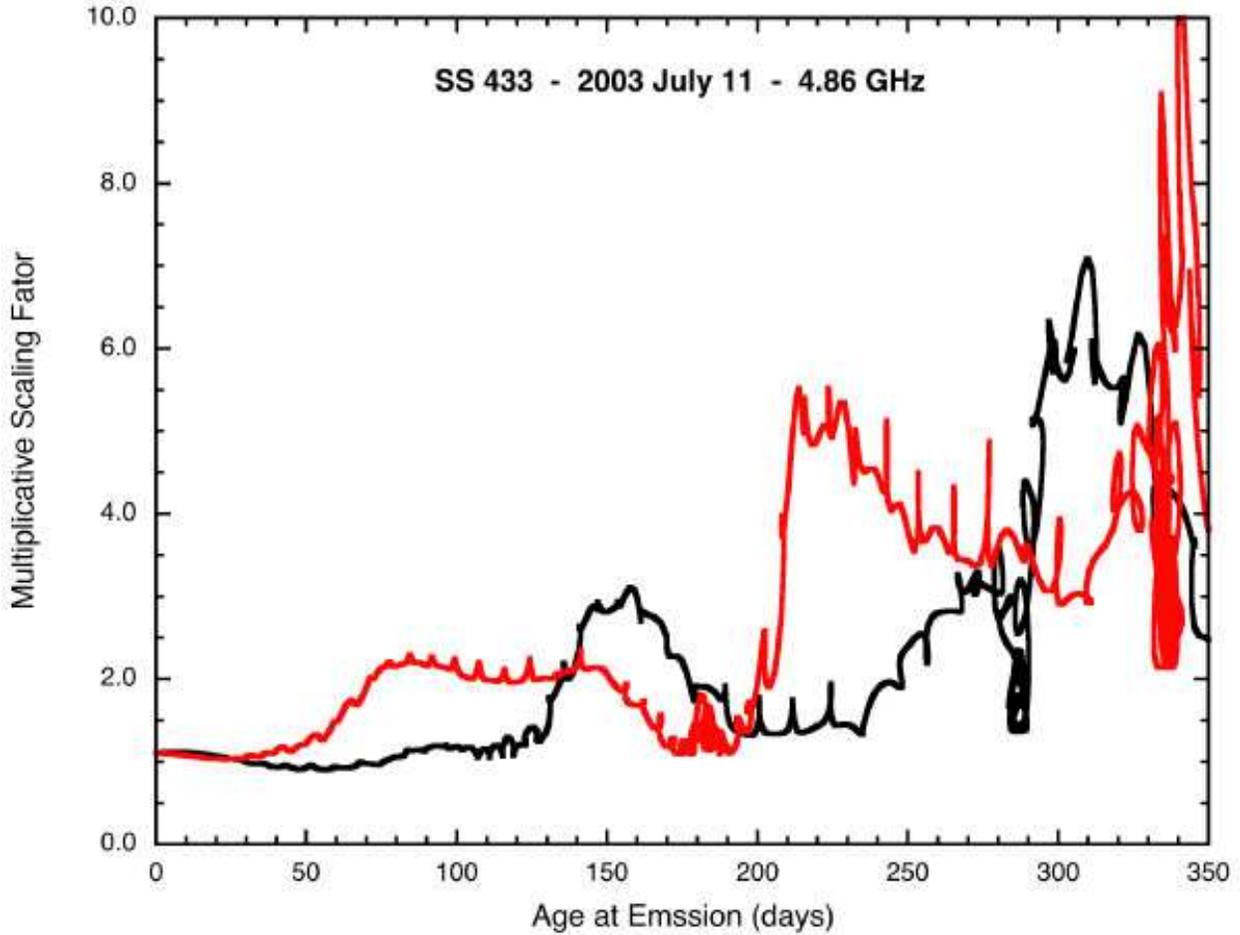}
\caption[]{Combined normalizations for the jets in SS\,433, expressed as multiplicative factors to be applied to observed total intensities, as functions of  the age of the jet material. These are the reciprocals of the beam-summed Doppler boosts for $n=2$. The black lines are for the east jet, the reds lines (grey in black \& white) for the west jet. Curves for $n=3$ are similar and not shown.
\label{fig:Norm2}}
\end{figure}

In order to study the intrinsic properties of the jets, we used the kinematic model to determine the effects on the observed jets of projection and Doppler beaming. In what follows, we will present jet properties as functions of age $\tau$ instead of the birth epoch $t$ because we expect the aging of the jet material to be the most important factor determining its properties. Figure~\ref{fig:LTT} shows $\tau$ for both east and west jets as functions of $t$, and it can be used to estimate $\tau$ at any location on the images, using the average proper motion of $\mu \simeq 8 \mbox{ mas d}^{-1}$. Projection effects were determined using model jets consisting of discrete components equally-spaced in birth epoch $t$; the projected density on the sky was determined by performing a beam-averaged count of the number of components per beam area as functions of position down the kinematic locus of the the two jets. This was normalized to unity at the core. The Doppler factor $D$ of each component was found from the kinematic model, and Doppler boosts calculated as $D^{\alpha+n}$, where a continuous jet has $n=2$ and an isolated component has $n=3$ (see Appendix~2 (\S\ref{s:app2})), and $\alpha \simeq 0.7$ is the spectral index of the jet ($F_\nu \propto \nu^{-\alpha}$;  Paper~III, Paper~IV). The total normalization factors were obtained by performing a beam-weighted sum of the boosts as a function of position down the model jets; all quantities were determined as functions of $\tau$. Figure~\ref{fig:Norm2} shows combined normalizations for the jets in SS\,433, expressed as multiplicative factors to be applied to observed total intensities, as functions of  $\tau$, for $n = 2$. Curves for $n=3$ are similar and not shown. Data from Figure~\ref{fig:Norm2} will be used below to create model jet intensity curves.  The ratio of the normalizations for the two values of $n$ ranges between 0.92  and 1.12, sufficiently different from unity to distinguish the two in the regions $\tau \leq 150~\mbox{ days}$ where the jet flux is greatest. 

\subsection{Determination of the Intrinsic Structure}
\label{s:IntrinsicStructure}

\begin{figure}[t]{}
\plotone{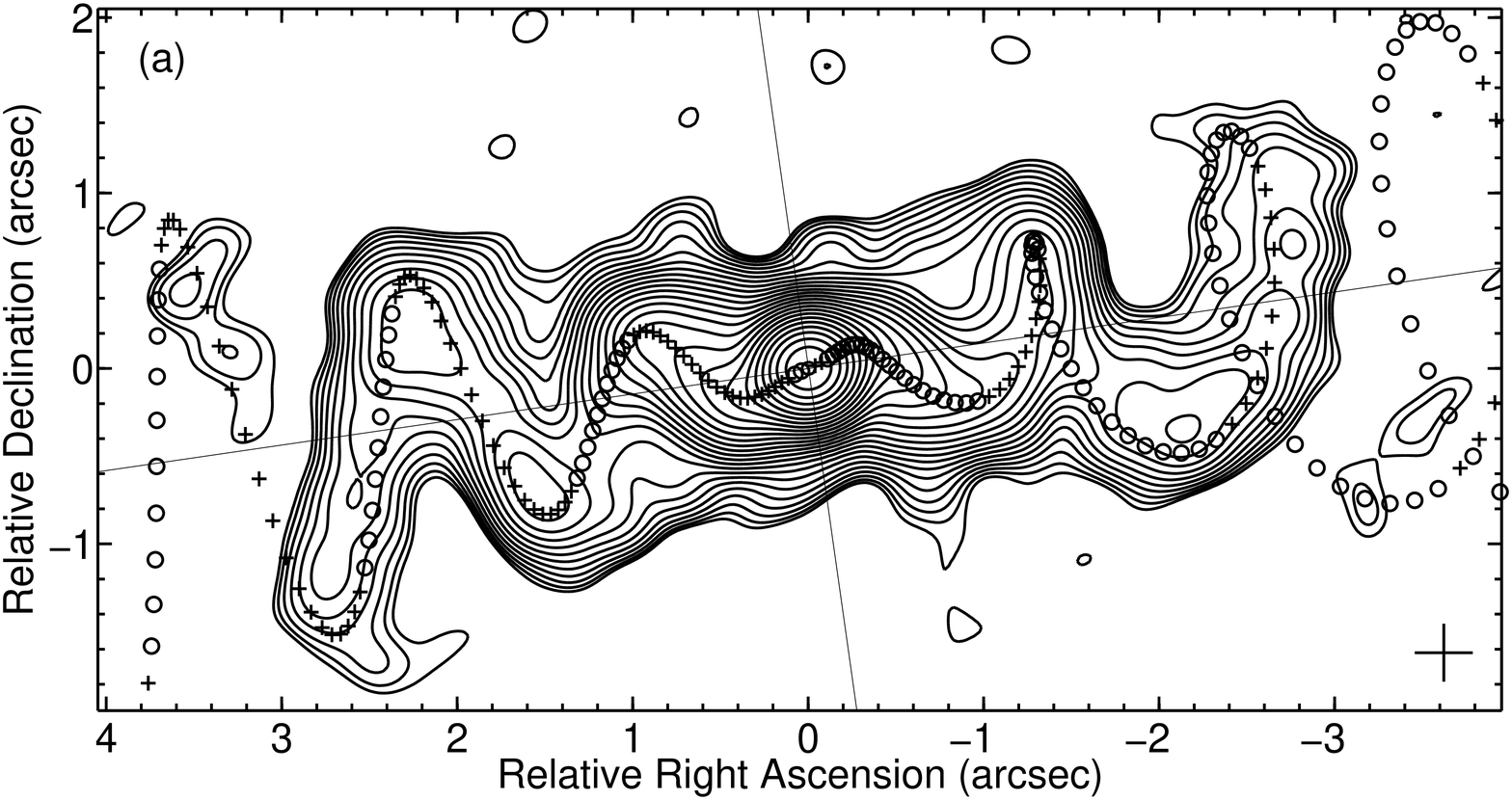}
\plotone{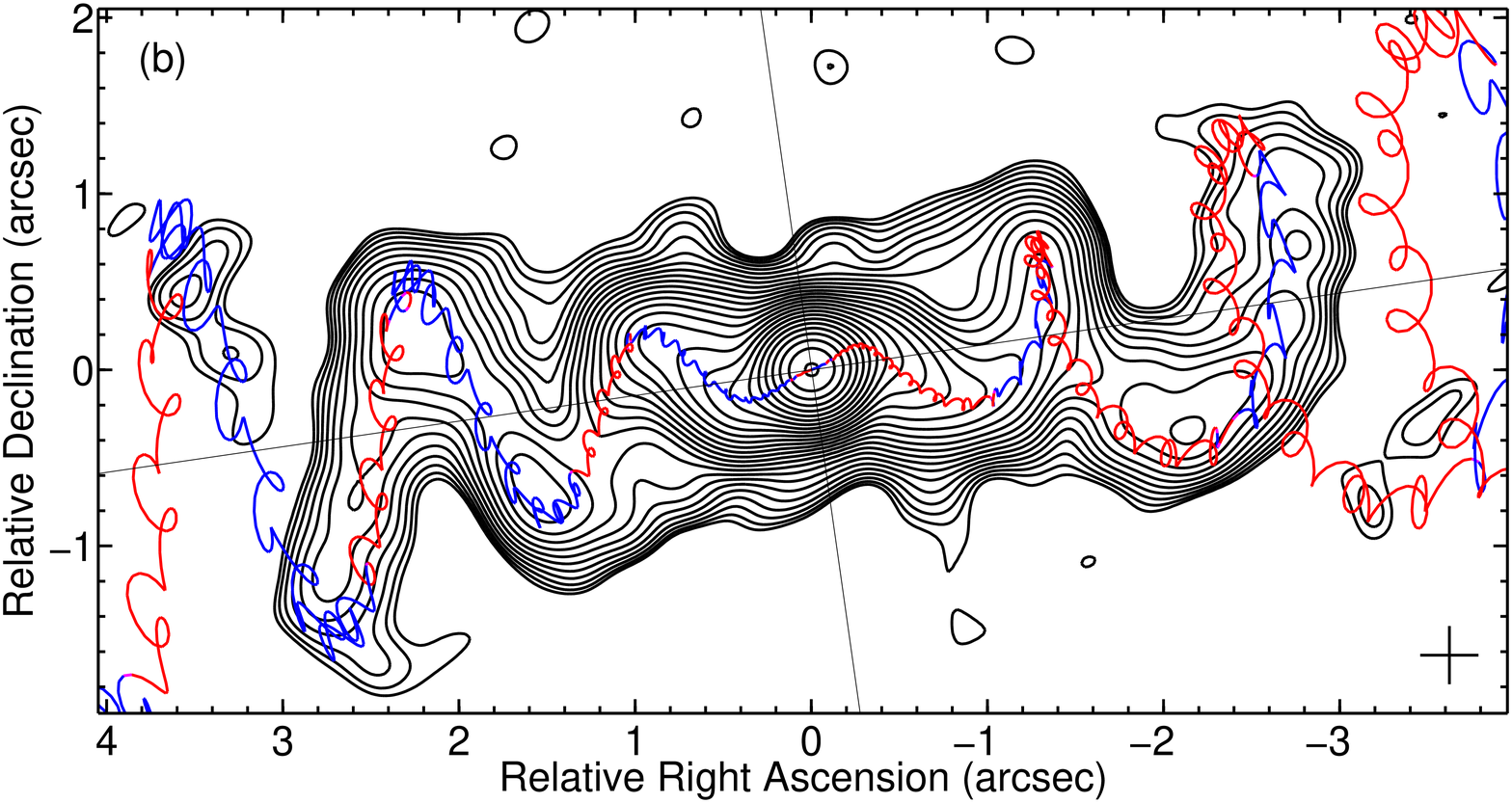}
\caption[]{Total intensity images of SS\,433 made from 4.86~GHz VLA A-array data collected 2003 July 11, using uniform weighting (ROBUST = $-5$). In (a) we show a contour image of SS\,433, with contours of total intensity $I$; the contours are spaced by factors of $\sqrt{2}$; the peak is 301~mJy ~beam$^{-1}$, the minimum contour is 50~$\mu$Jy~beam$^{-1}$, and the RMS noise is 21~$\mu$Jy~beam$^{-1}$. The kinematic model without  jet velocity variations or nutation is shown as pluses for oncoming material and circles for retreating material, with components emitted at 5~day intervals. (b) Same as (a), except that the kinematic model including jet velocity variations and nutation is shown as  blue and red lines, where these are oncoming and retreating parts, respectively. In these images the beam is $0.33\arcsec$ by $0.32\arcsec$. The scale of the figures is such that $1\, \arcsec = 5500\, \mbox{AU}$, or 125~days of mean proper motion of 8~mas day$^{-1}$.
\label{fig:CUN}}
\end{figure}

\begin{figure}[t]
\plotone{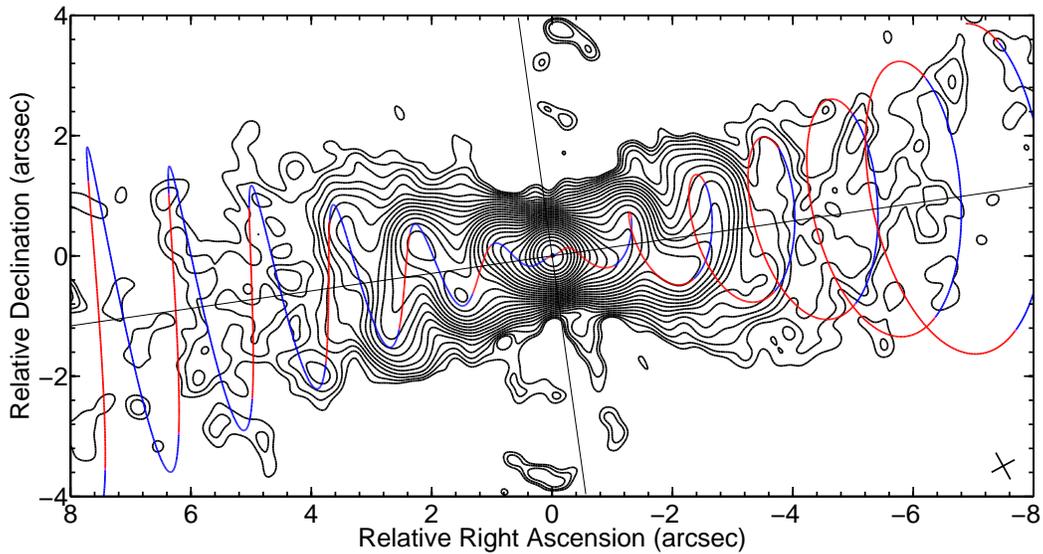}
\caption[]{Total intensity image of SS\,433 made from 4.86~GHz VLA A-array data collected 2003 July 11, using natural weighting (ROBUST = +5). We show contours of total intensity $I$; the contours are spaced by factors of $\sqrt{2}$, the peak is 302~mJy ~beam$^{-1}$, and the minimum contour is 26~$\mu$Jy~beam$^{-1}$, twice the  root-mean-square noise of 13~$\mu$Jy~beam$^{-1}$. The kinematic model (not including jet velocity variations and nutation) is shown as blue and red lines, where these are oncoming and retreating parts, respectively. The beam is shown as a cross, and has full width at half maximum of $0.47\arcsec$.
\label{fig:CNA}}
\end{figure}

\begin{figure}[t]
\plottwo{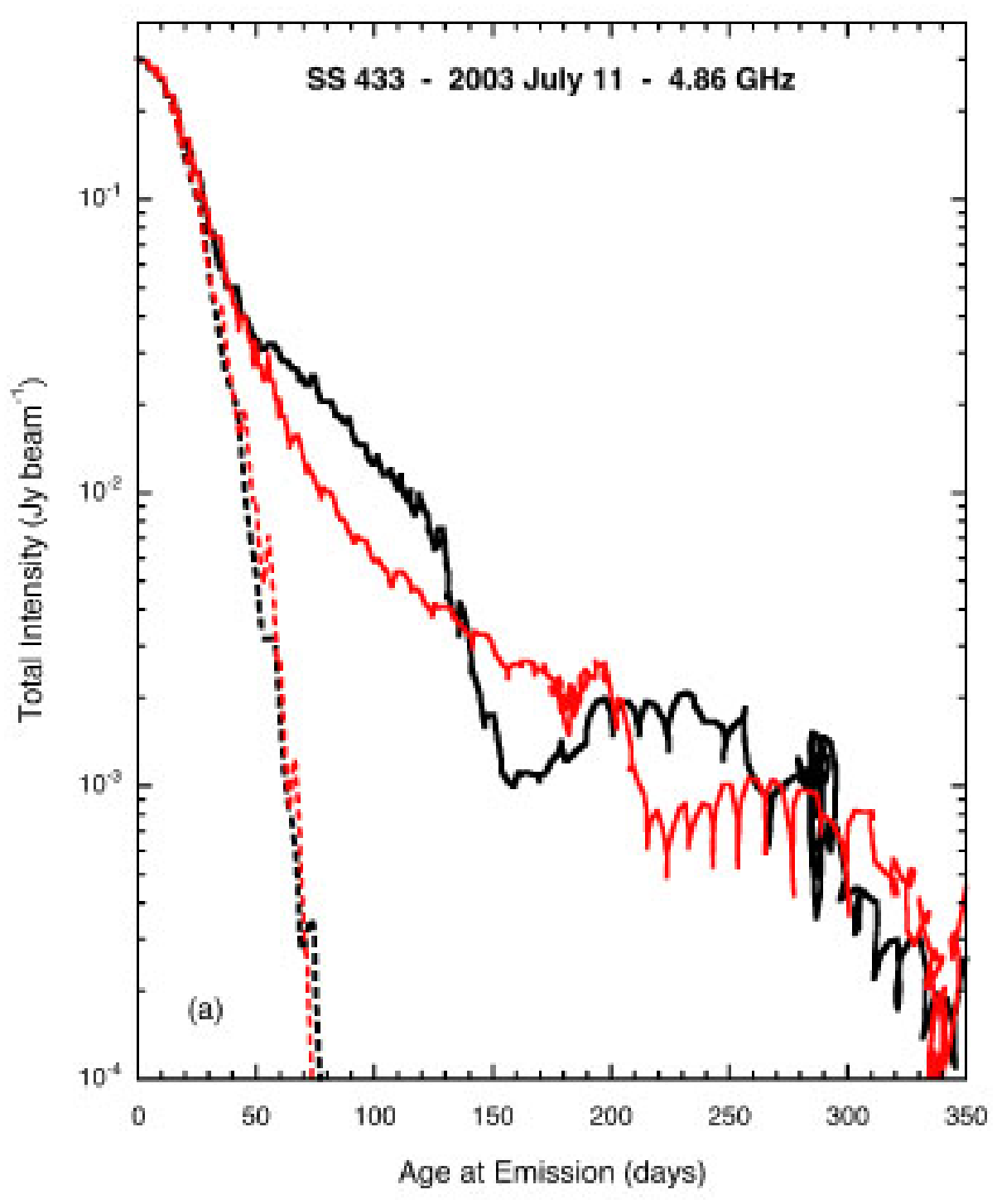}{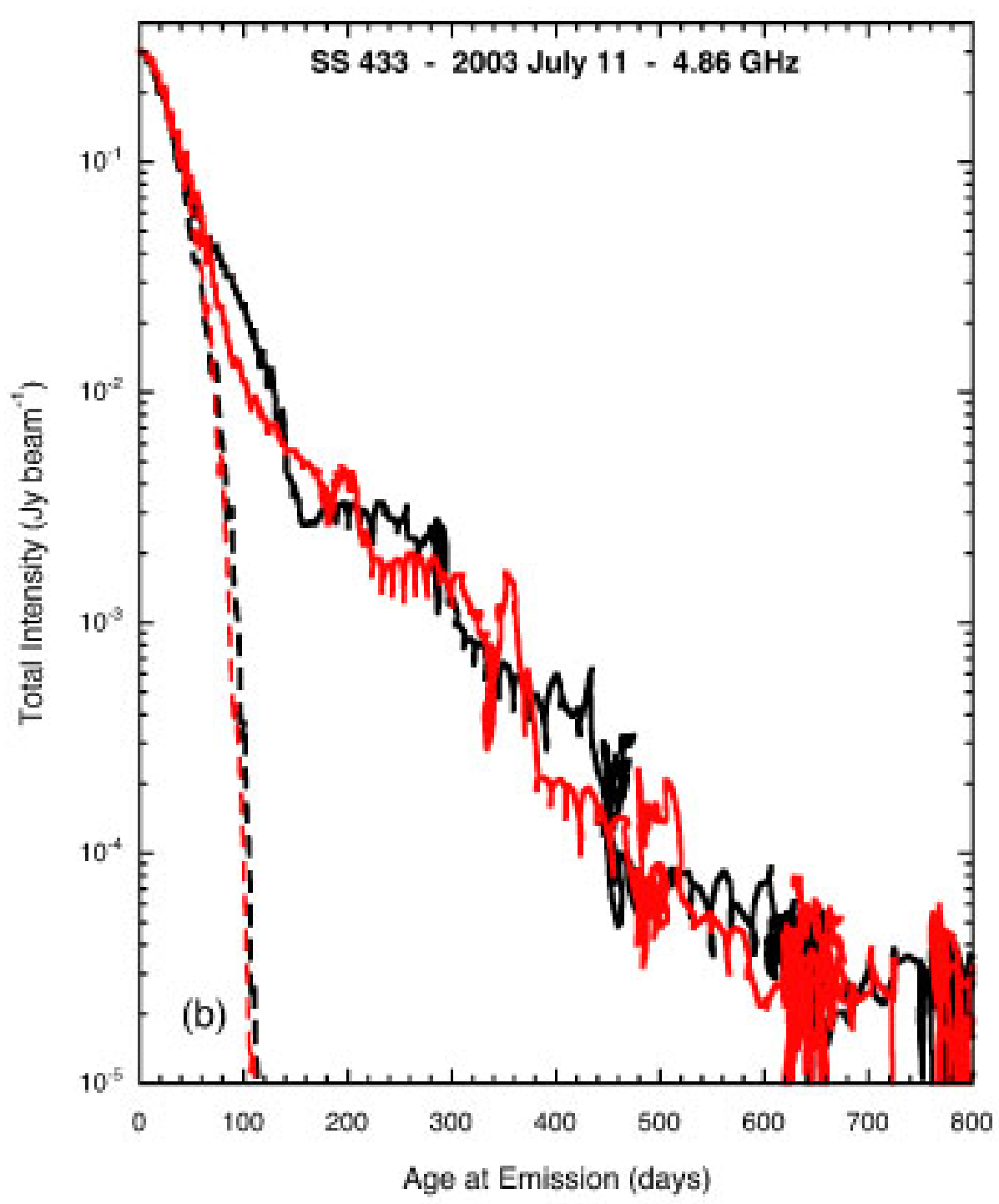}
\caption[]{Total intensity profiles of the jets of SS\,433 as a function of the age of the jet material derived from (a) the image in Figure~\ref{fig:CUN}; the root-mean-square noise is $21 \, \mu$Jy/beam, and (b) derived from the image in Figure~\ref{fig:CNA}; the root-mean-square noise is $13 \, \mu$Jy/beam. The black lines are for the east jet, the red ones (grey in black \& white) for the west jet, and the broken lines show the core contributions to the total intensity. Jet emission is detectable out to ages at emission of at least 800~days ($\sim 6 \arcsec$ from the core) in both jets.
\label{fig:FUN(age)}}
\end{figure}

Figure~\ref{fig:FUN(age)} shows the profiles down the model locus of the observed total intensities of the jets derived from the image in Figure~\ref{fig:CUN} and the naturally-weighted image in Figure~\ref{fig:CNA}, as functions of $\tau$. For the region $\tau \ge 60 \mbox{ days}$, contamination by the core is less than 1\% of the jet intensity in Figure~\ref{fig:CUN}; the beam occupies the first 75 days in Figure~\ref{fig:CNA}. The loops in the curves for both jets are the result of the fact that material of multiple different birth epochs can have the same age as a result of differing light travel times. 

It is useful to know that the places along each jet where the radial velocity switches sign and the jet motions are both in the plane of the sky are located at $t  = \tau = 14, \,125,\,176,\, 288,\, 338 ,\,450, \mbox{and} \, 500 $~days. At these ages the projection factors and Doppler boots of the two jets are equal, so if  the two jets are intrinsically the same, the raw total intensities should be the same. Examination of the intensity curves shows that, allowing for the beams of full-width at half-maximum about $\Delta \tau \simeq 40 \mbox{ days}$ for Figure~\ref{fig:FUN(age)}(a) and $\simeq 60 \mbox{ days}$ for Figure~\ref{fig:FUN(age)}(b), this is satisfied.

\begin{figure}
\plottwo{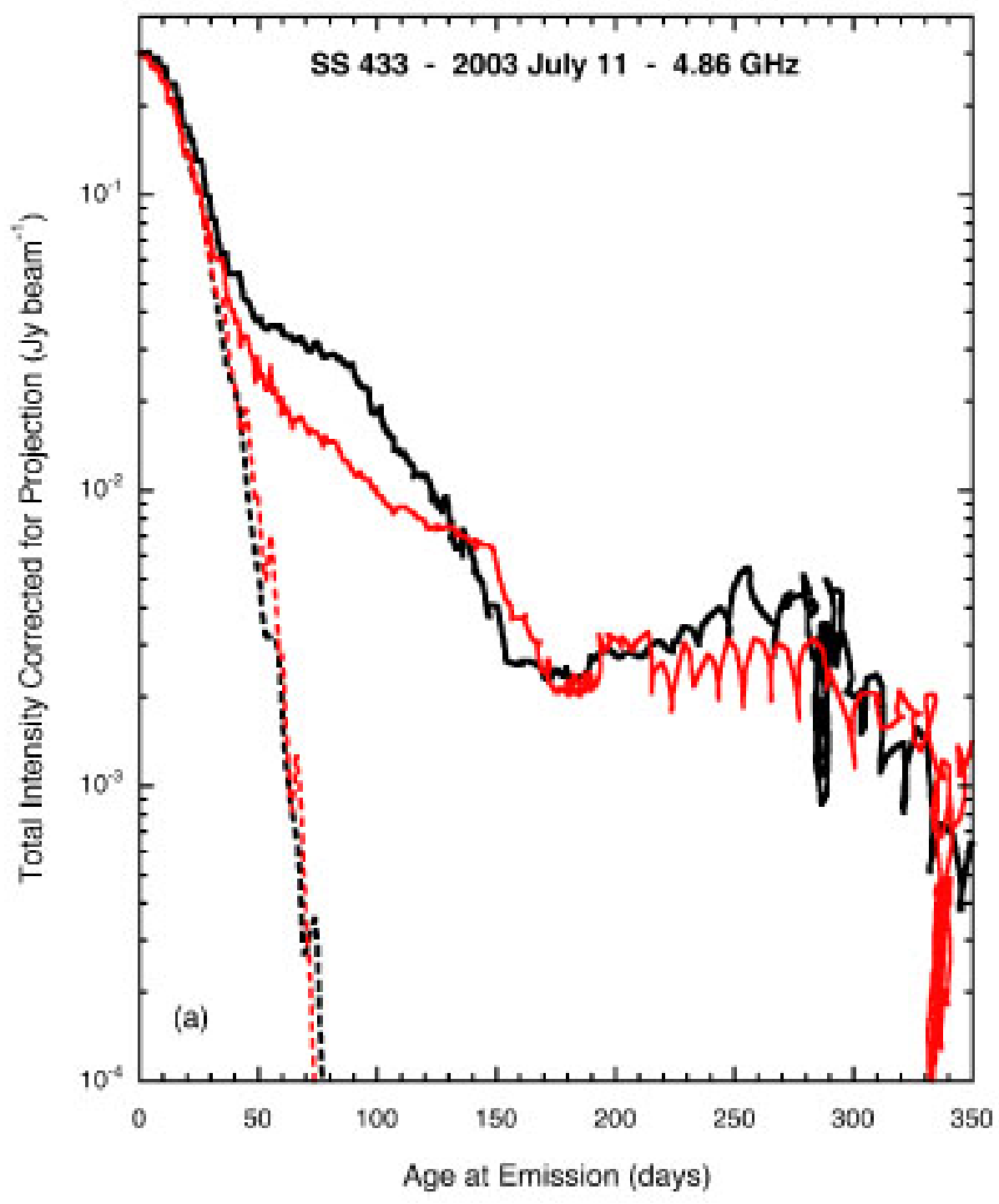}{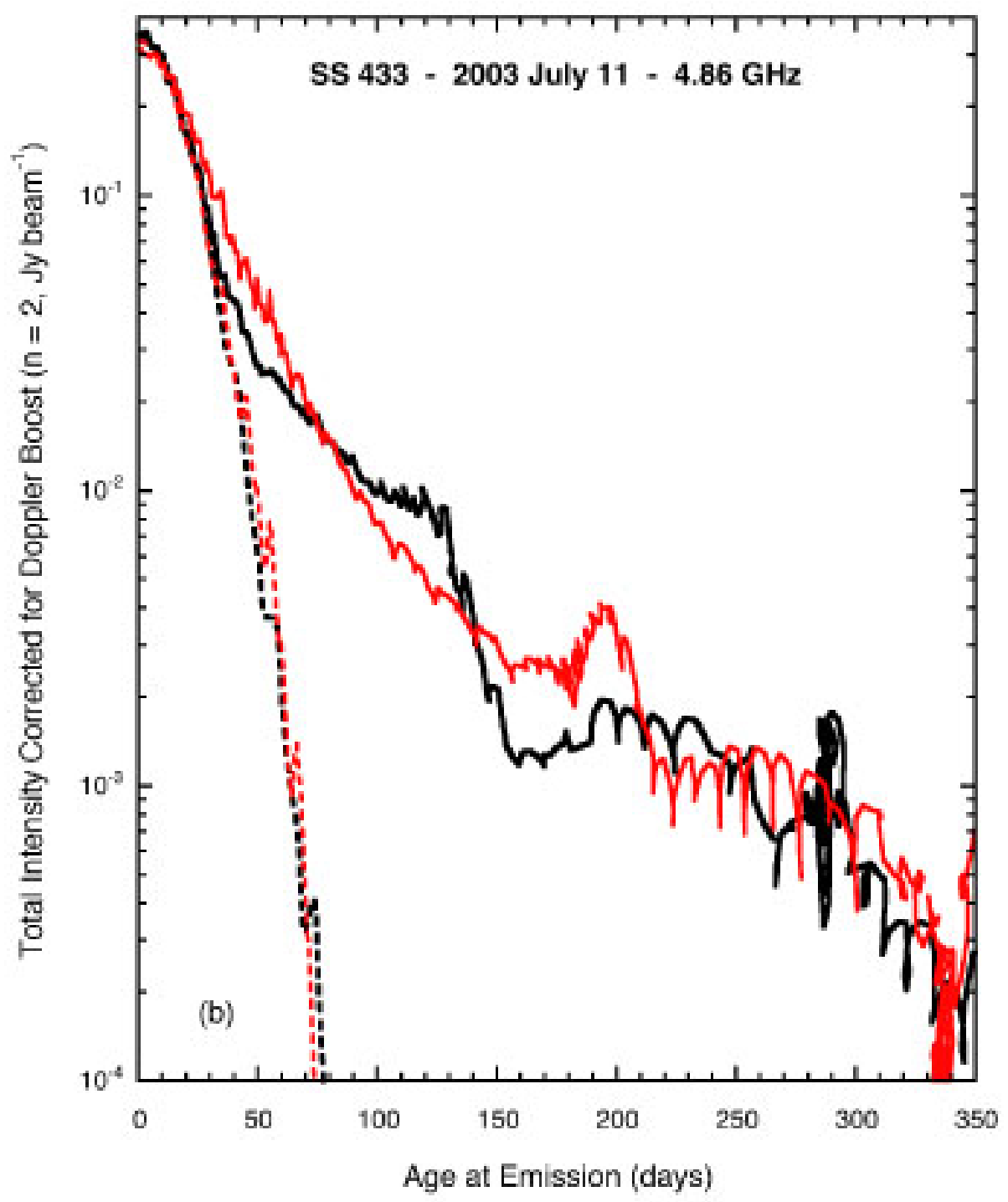}
\caption[]{Total intensity profiles normalized for (a) projection effects alone, and (b) for Doppler boosts ($n=2$) alone (very similar to $n=3$, not shown). The black lines are for the east jet, the red lines (grey in black \& white) for the west jet, and the broken lines show the core contributions.
\label{fig:F1F2(age)}}
\end{figure}

\begin{figure}[t] 
\plottwo{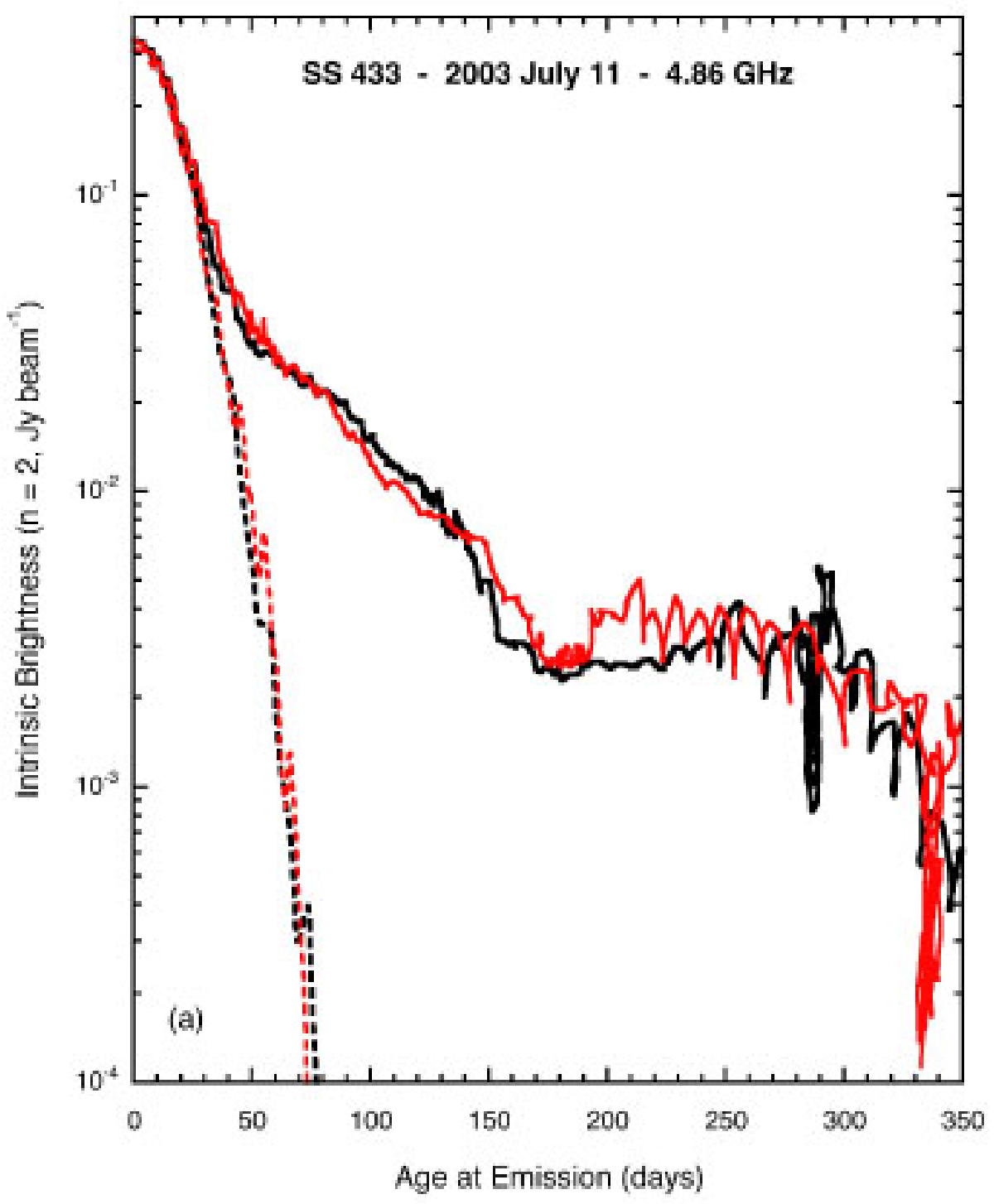}{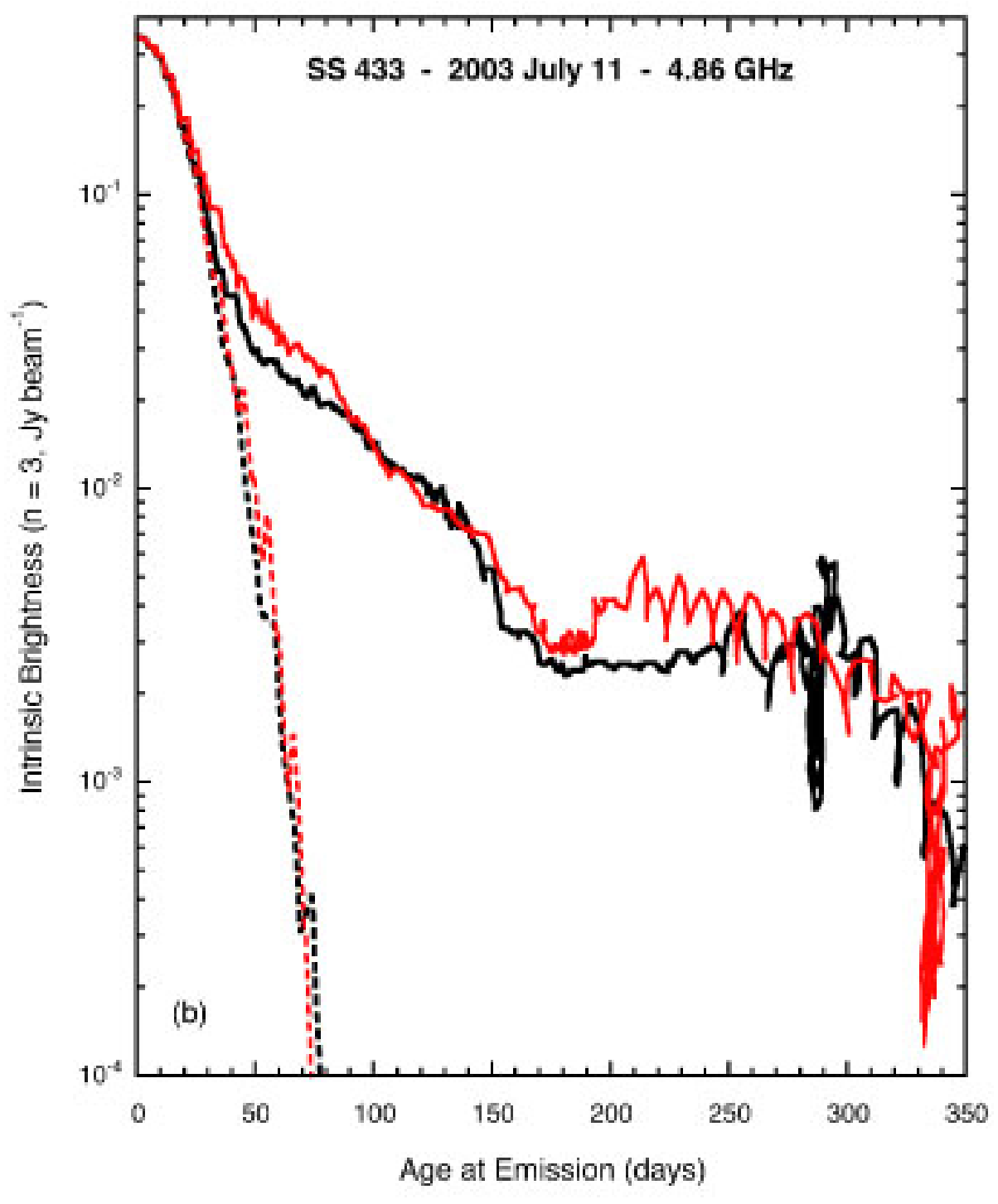}
\caption[]{Intrinsic brightness profiles of the jets of SS\,433 for (a) $n = 2$ and (b) for $n=3$ as functions of the age of the jet material, derived from the  image in Figure~\ref{fig:CUN}. The black lines are for the east jet, the red lines (grey in black \& white) are for the west jet, and the broken lines show the core contributions to the intrinsic brightness. 
\label{fig:NormCUN}}
\end{figure}

As can be seen from Figure~\ref{fig:F1F2(age)}, neither normalization for projection effects alone nor for Doppler boosting alone results in east and west jets appearing the same. This is true for either choice of $n$. Figure~\ref{fig:NormCUN} shows the intrinsic brightness\footnote{We use this term to refer to the total intensities that we would observe were we in a frame co-moving with a piece of the jet.} profiles of the jets in SS\,433 created by normalizing for both projection effects and Doppler boosting for $n=2$ and $n=3$, as functions of $\tau$, derived from the image in Figure~\ref{fig:CUN}.  Especially in the case $n=2$, the intrinsic brightnesses of the two jets are very similar when compared for equal ages. 
The small discrepancies between the derived intrinsic brightnesses of the east and west jets for $n=2$ in the range $80 \leq \tau \leq 220\mbox{ days}$ can be understood as artifacts due to the tight bend in the east jet at RA~$\simeq 1^{\arcsec}$ and the first loop in the west jet,  RA~$\simeq -1.2^{\arcsec}$; these correspond to ages  $80 \leq \tau \leq 130\mbox{ days}$ and  $140 \leq \tau \leq 220\mbox{ days}$, respectively. In the first range the measured total intensity for the east jet is artificially high, as seen in Fig.~\ref{fig:NormCUN}(a), due to two parts of the jet locus being ``in the beam'' at the same time. In the second range, the total intensity of the west jet is falsely measured to be high, again corresponding to Fig.~\ref{fig:NormCUN}(a). The normalization process cannot account for the contamination of one side of the loop by the part of the other that is in the same beam (the procedure lacks {\em a priori} information about the ratio of the two contributions).  This is minimal at the cusp of the loop, $\tau \simeq 180\mbox{ days}$. However, when we reconstruct the west jet from the intrinsic brightness of the east jet and the normalization model  this is accounted for, and the artifacts largely disappear (see below). While the same arguments apply to $n=3$, they cannot account for the discrepancy over $60 \leq \tau \leq 90 \mbox{ days}$; the agreement for $n=3$ is simply not as good where the jet is brightest, suggesting that $n=2$. 

\begin{figure}[t]
\plotone{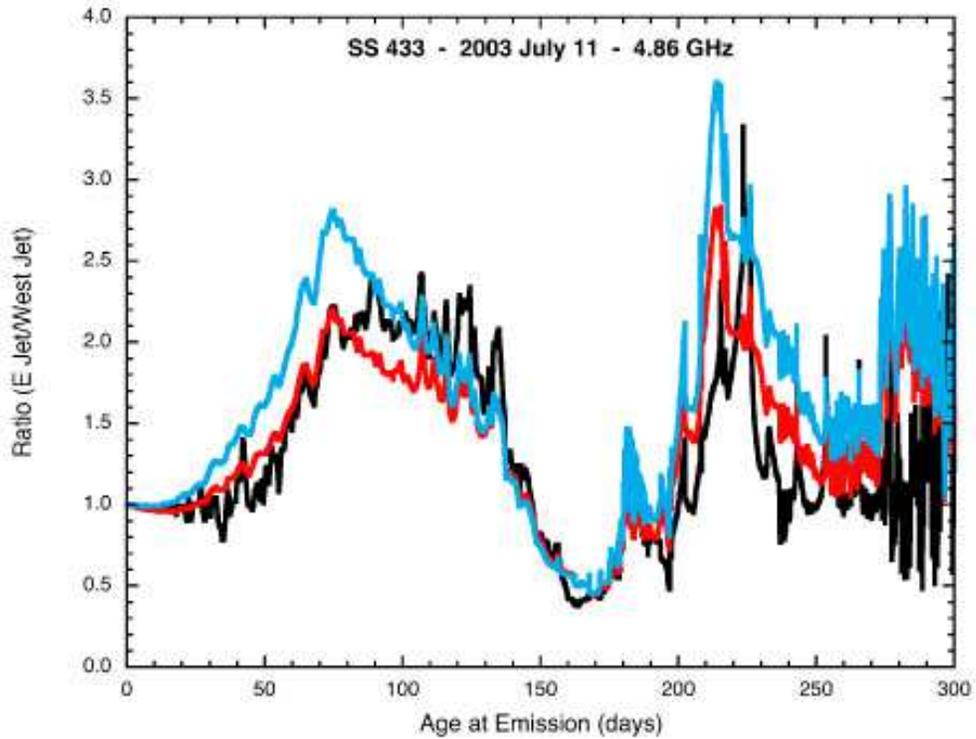}
\caption[]{Observed total intensity ratio and normalization factor ratios, both in the form east jet divided by west jet, as functions of the age of the jet material $\tau$.  The black line is the observed total intensity ratio, and the red line (grey in B\&W) is the ratio of multiplicative normalization factors for $n=2$ (Fig.~\ref{fig:Norm2}), the blue curve for $n=3$, which  provides a less satisfactory fit. If the two jets are identical and our normalization is correct, the red or blue curve should lie on top of the black curve.
\label{fig:MasterRatiosPlot}}
\end{figure}

We can see the results of the normalizations in another way if we examine the total intensity ratio of the two jets. Figure~\ref{fig:MasterRatiosPlot} shows the ratio of observed total intensities in the form east jet divided by west jet and the predicted ratios for complete correction for $n=2$ and $n=3$, all as functions of $\tau$. The fit of the combined normalization to the data is quite good for $n=2$; the prediction for $n=3$ is not as good a fit. In fact, the discrepancies between the data ratio and the $n=2$ model ratio are explained in the same way described above. In the range $80 \leq \tau \leq 130\mbox{ days}$, the east jet is measured to be artificially high, raising the measured ratio above the model ratio, in agreement with the figure; in the range $140 \leq \tau \leq 220 \mbox{ days}$, the west jet is measured artificially high, producing the opposite effect, again in accord with the figure.

\begin{figure}[t]
\plottwo{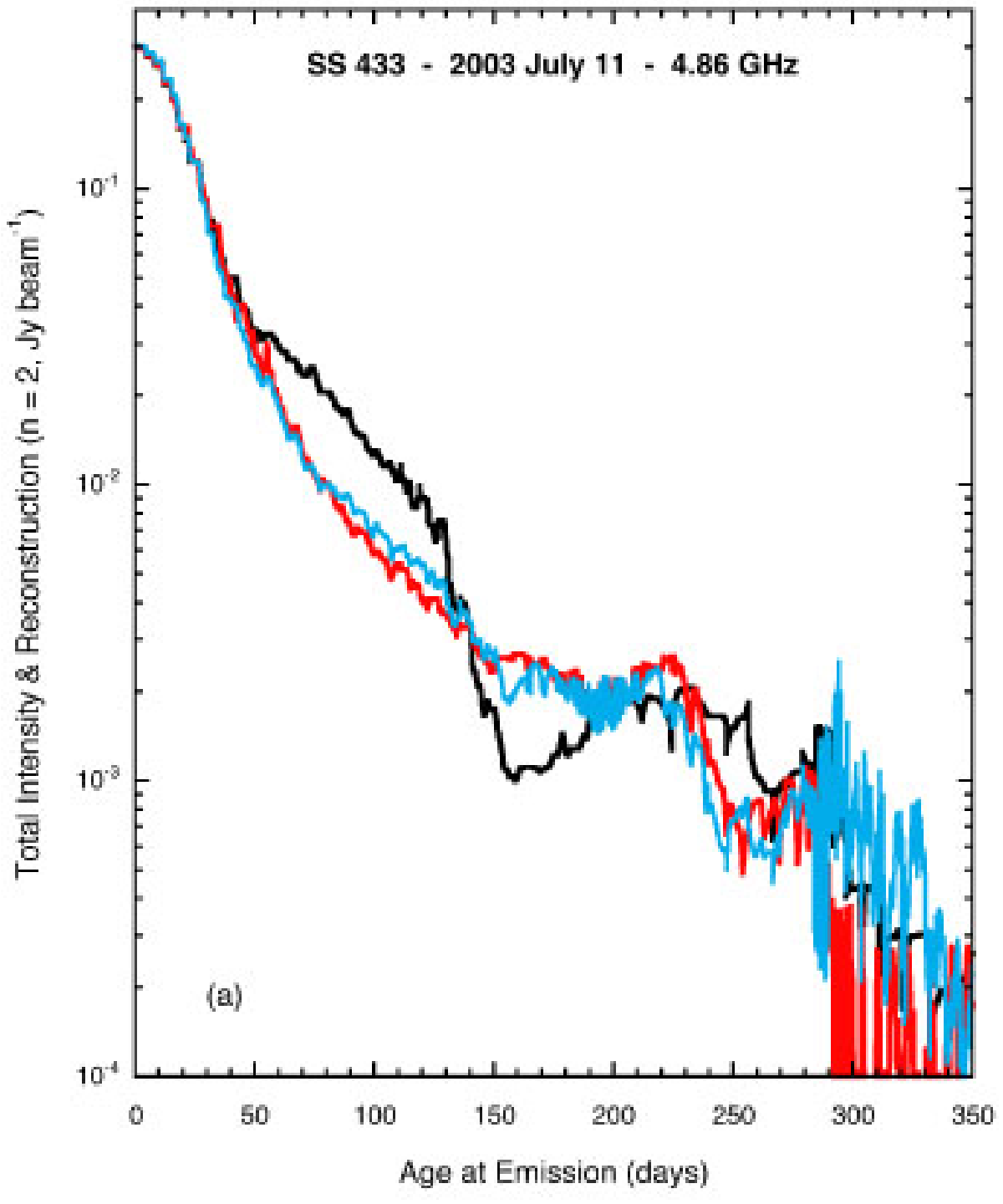}{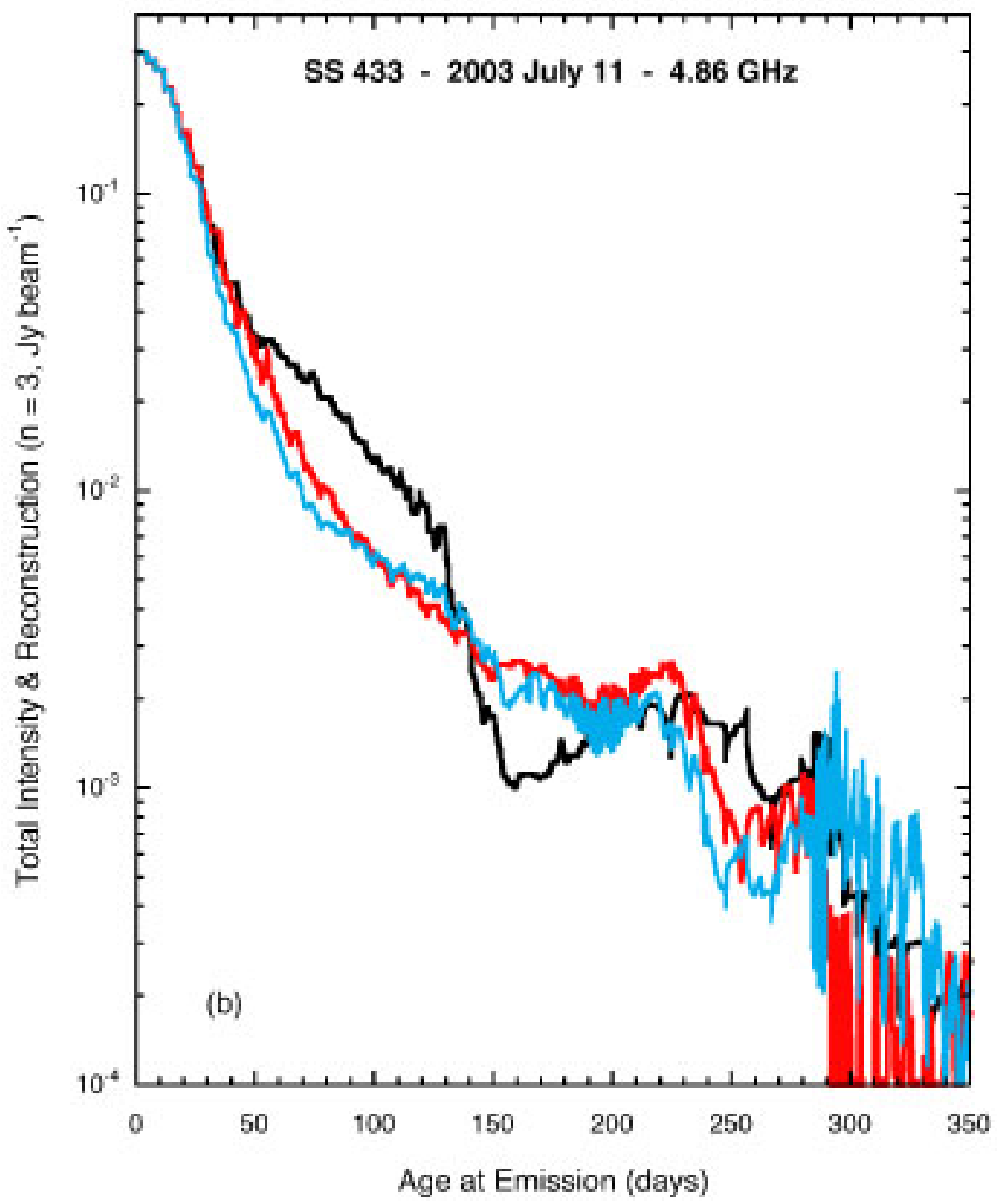}
\caption[]{Reconstruction of the west jet. We show the observed total intensities of east and west jets (black lines and red lines, respectively), and reconstructed total intensity profiles of the west jet (blue lines) as functions of the age of the jet material, for (a) $n=2$ and (b) for $n=3$. The data were obtained from the image in Figure~\ref{fig:CUN}. 
\label{fig:Recon1}}
\end{figure}

We can test the similarity of the two jets in a third way, by using the derived intrinsic brightness of the east jet in Figure~\ref{fig:NormCUN} and the model normalization factors in Figure~\ref{fig:Norm2} to predict the observed total intensity of the west jet. Comparisons for $n=2$ and $n=3$ are shown in Figure~\ref{fig:Recon1} for ages out to 350~days. In the region $\tau \leq 350$~days the east jet is a single-valued function of age, so it can be modeled (normalized) uniquely, and the loops in the west jet reconstructed free of artifacts. The small uncertainties in the intrinsic brightness  
of the east jet described above produce small disagreements over $60 \leq \tau \leq 200\mbox{ days}$, as expected.
The overall fit is better for $n=2$ than for $n=3$, also in agreement with the results above. Because $t$ becomes a multiply-valued function of $\tau$ beyond ages of about $\tau = 350$~days, it is not possible to reconstruct the west jet uniquely beyond this point. 

Finally, if we regard $n$ simply as a phenomenological parameter, examination of Figs.~\ref{fig:NormCUN}--\ref{fig:Recon1} shows that $n=2$ is favored over $n=3$ because agreement of the intrinsic brightness of the two jets where they are brightest is superior. 

In summary, we conclude that {\em in SS\,433 jets the observed total intensities are compatible with the twin jets being intrinsically identical, and behaving as continuous jets.} We address the question of the expected behavior of the observed total intensity on the Doppler boosts in Appendix~2 (\S\ref{s:app2}).


\section{DISCUSSION}
\label{s:Discussion}

\begin{figure}[t]
\plottwo{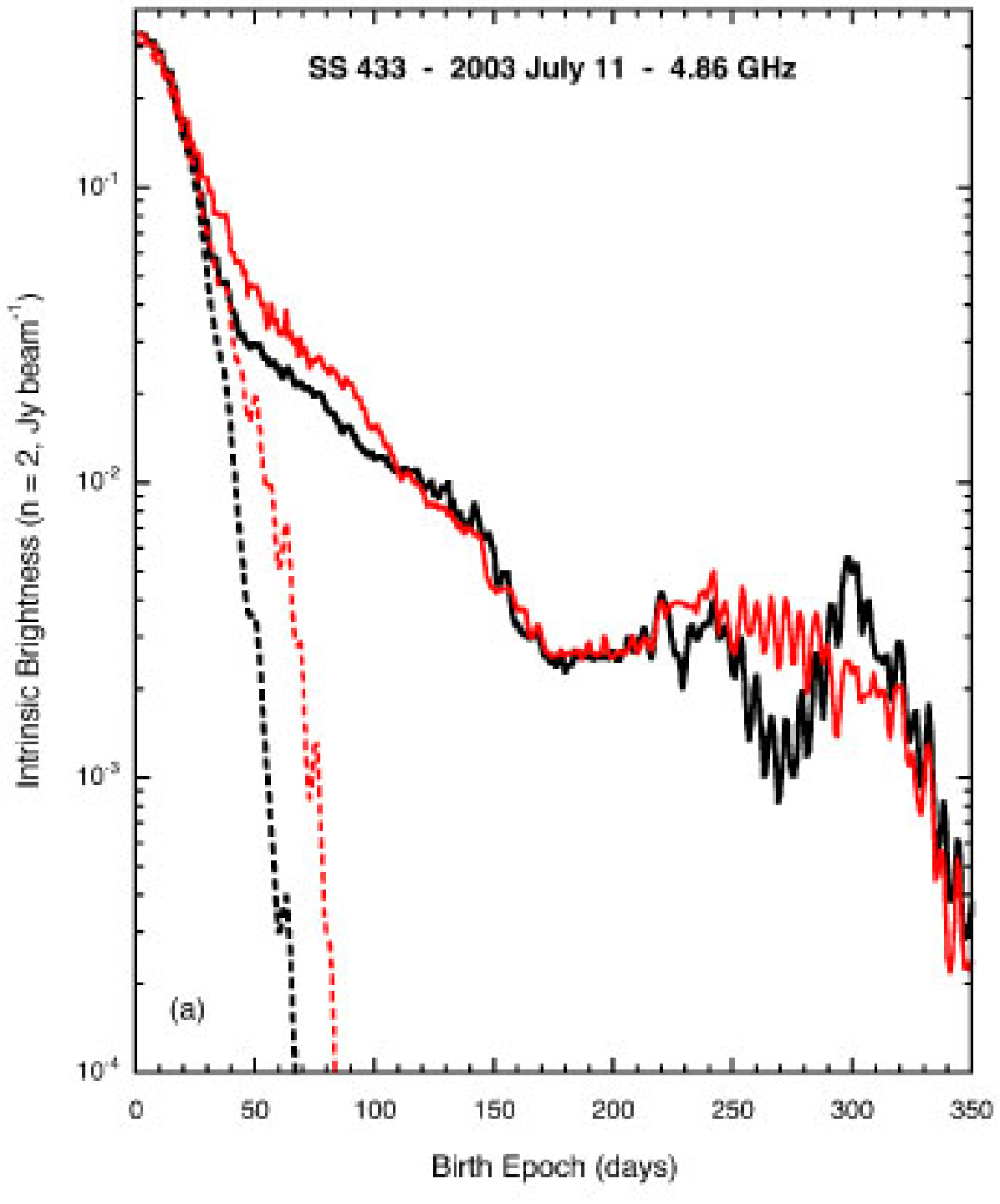}{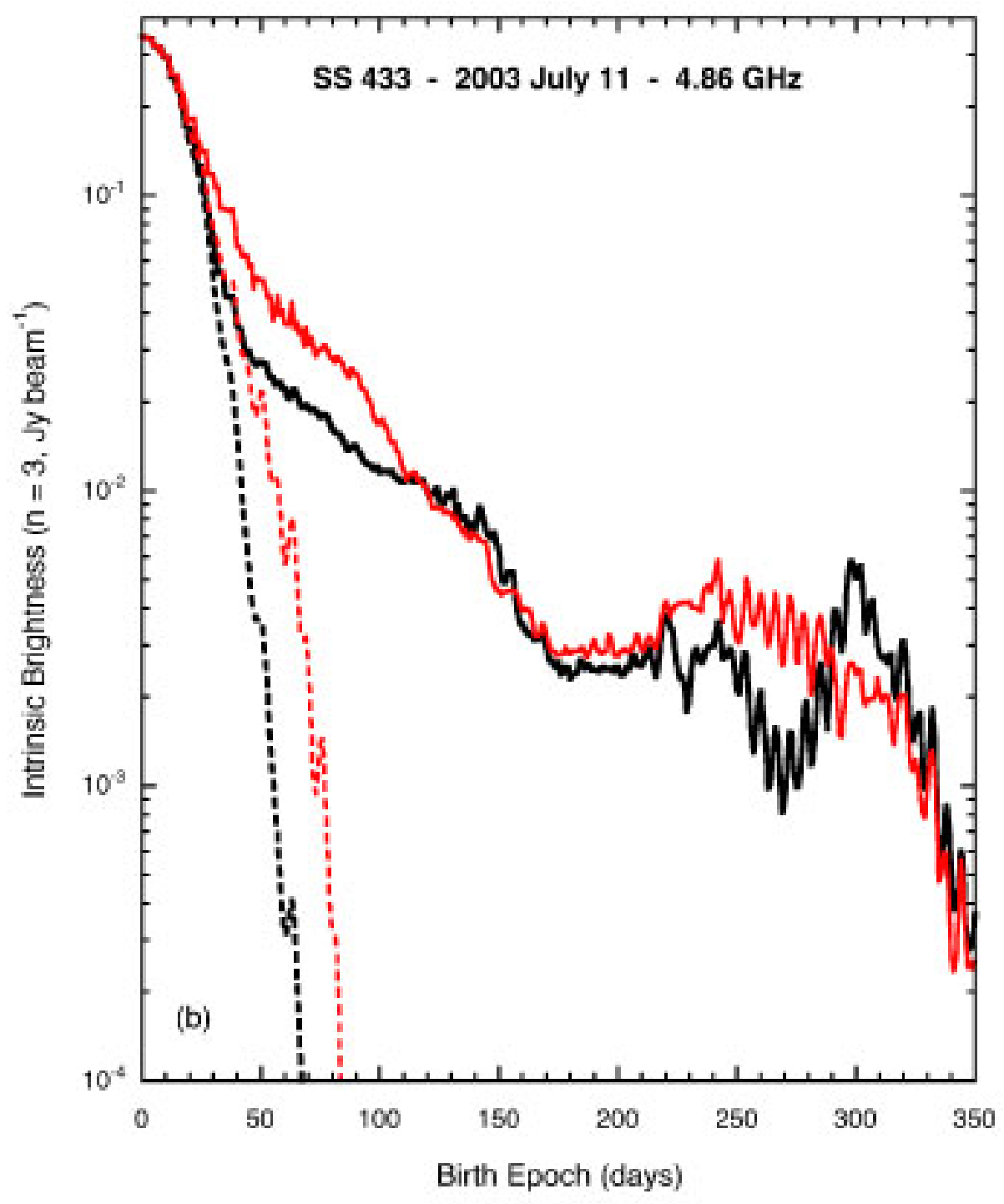}
\caption[]{Intrinsic brightness profiles of the jets of SS\,433 for (a) $n=2$ and (b) for $n=3$ as functions of birth epoch, derived from the image in Figure~\ref{fig:CUN}. The black lines are for the east jet, the red lines (grey in black \& white) are for the west jet, and the broken lines show the core contributions to the brightness.
\label{fig:F12(t)}}
\end{figure}

Three interesting questions about the jets are: (i) do they contain features, such as breaks or peaks in the intensity profiles in either or both jets that are not the effect of projection and Doppler boosting on an otherwise smooth intensity distribution, (ii) if so, are they the same in the two jets, and (iii) if so, how do they behave in time. Such features might arise at the ejection of the jets (the core after all is quite variable; Paper~III), or as functions of their aging and/or propagation. Figure~\ref{fig:F12(t)} shows the intrinsic brightness for $n=2$ and $n=3$ as a function of birth epoch. Because this compares east and west jet material of different ages there is no {\em a priori} reason to expect matching features.  However, the regions of roughly constant brightness located over the range $150 \leq t \leq 250 \mbox{ days}$ in both jets  may be such features because we see the same behavior over the entire summers of 2003 (Paper III) and 2007 (Paper~IV), but at different age ranges. Thus the flattening of the decay curve may be the result of variations in the central engine. Comparison of the two jets is hindered by our inability to extract unique properties of the west jet through the first tight loop. Nonetheless, what does seem clear is that after the initial rapid falloff of each jet, there is a leveling-off, and then a somewhat slower decline with large observational uncertainties. 

\begin{figure}[t]
\plottwo{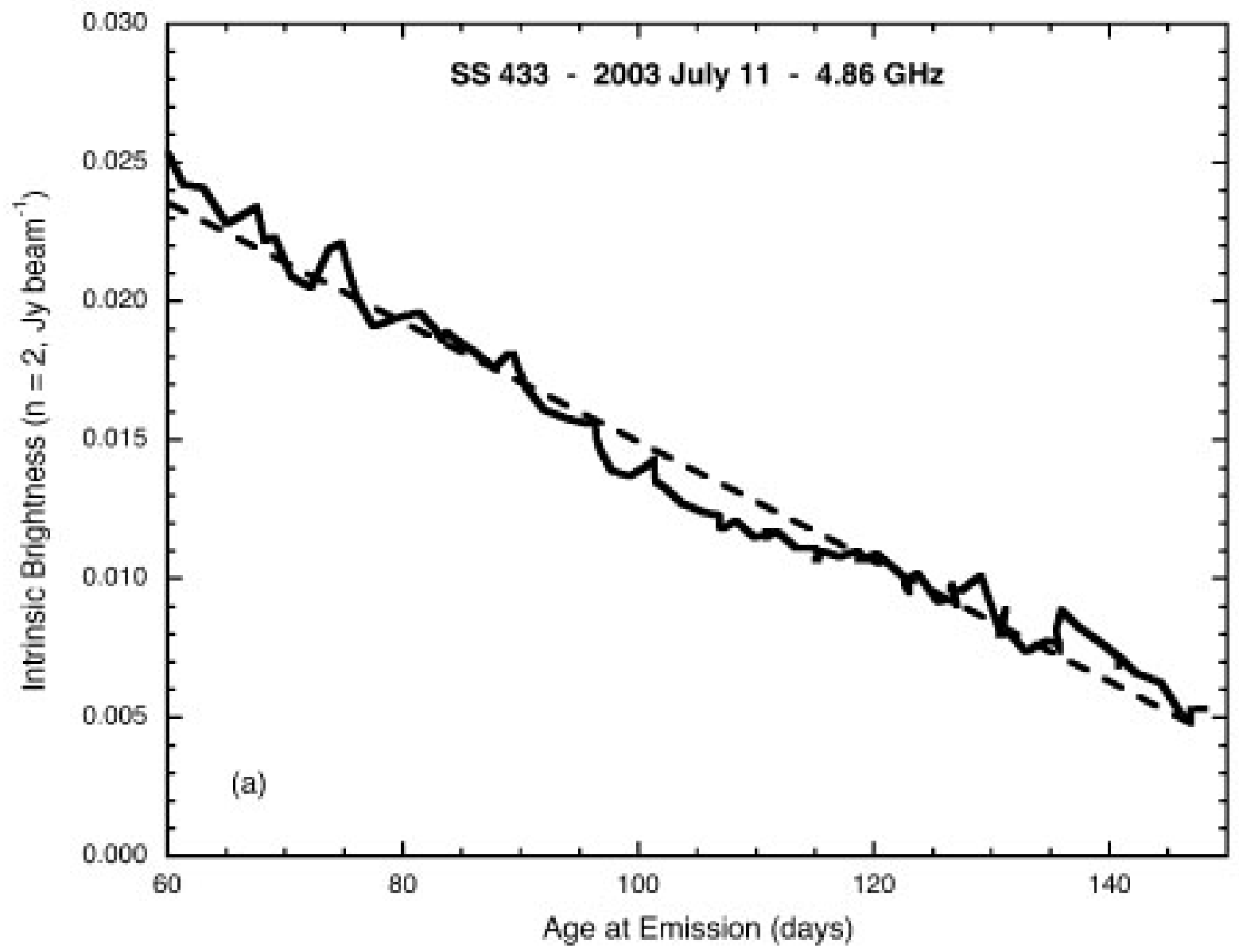}{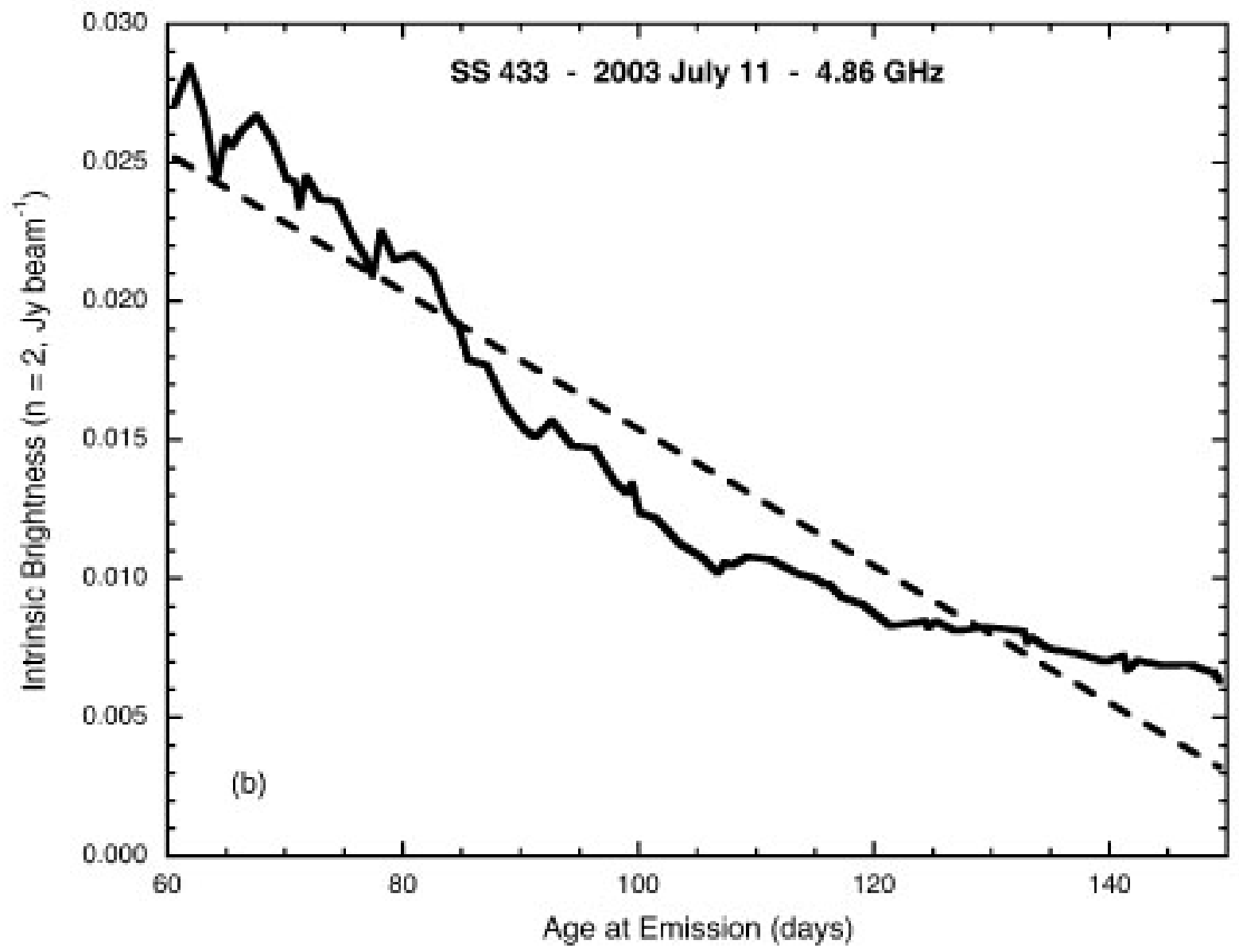}
\plottwo{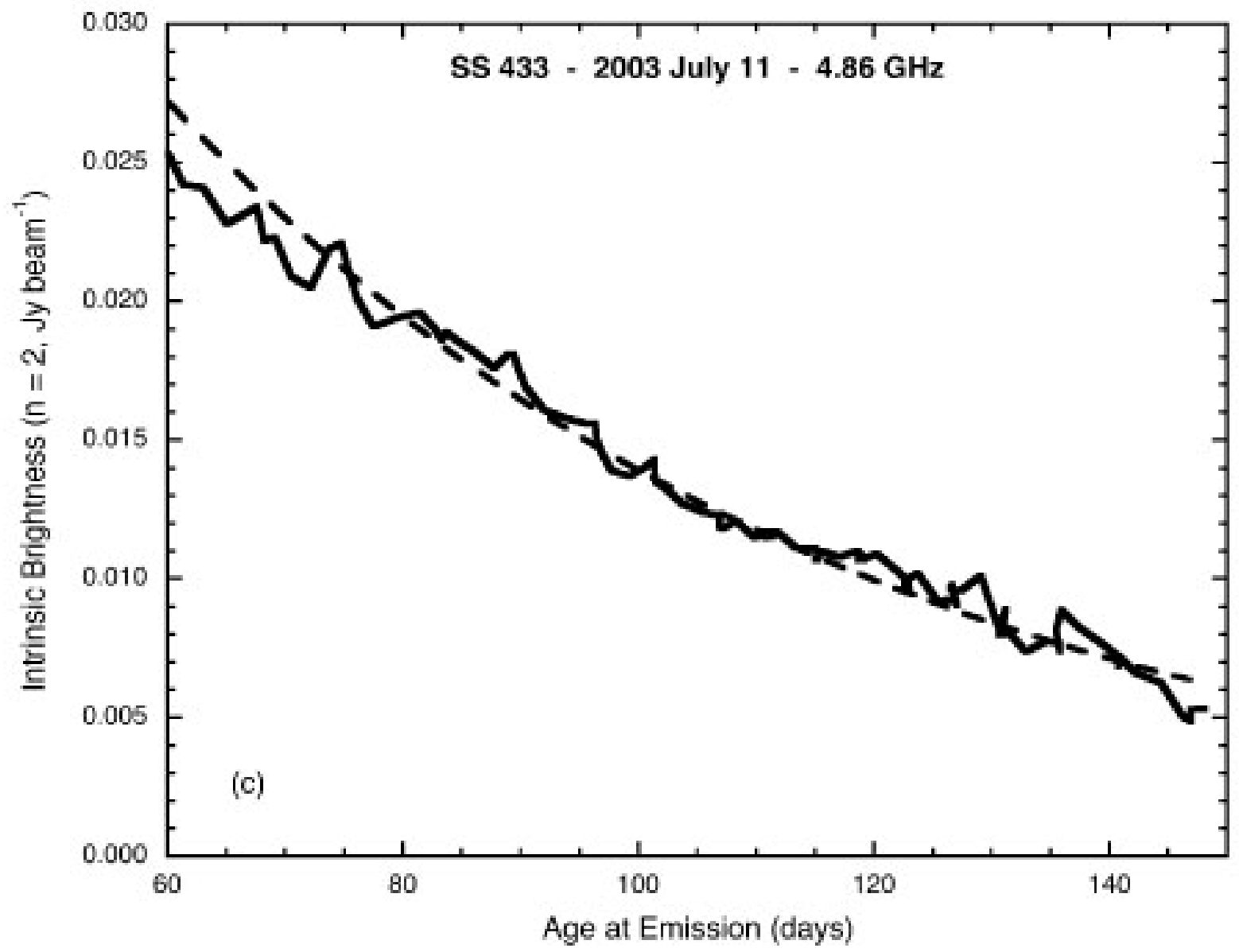}{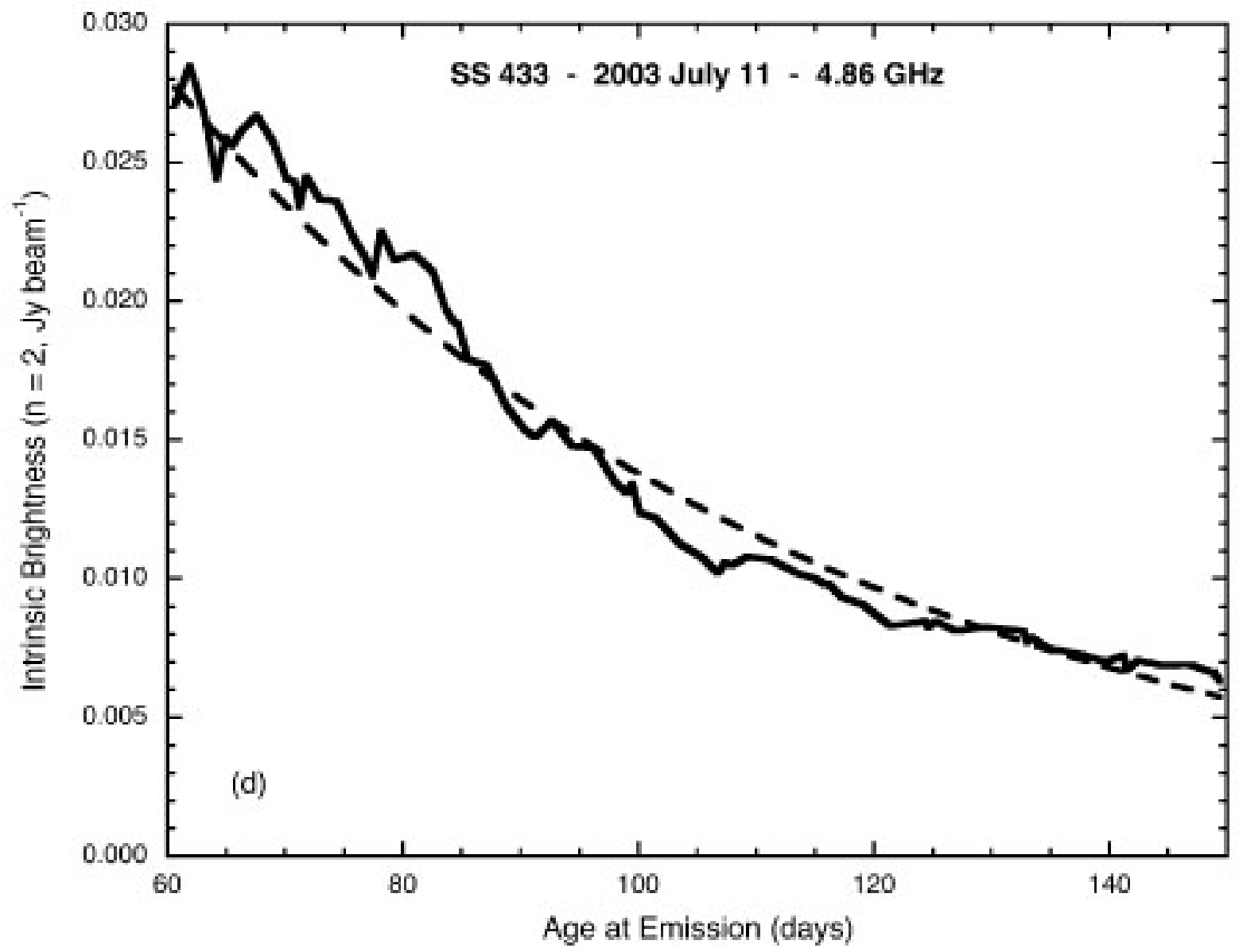}
\plottwo{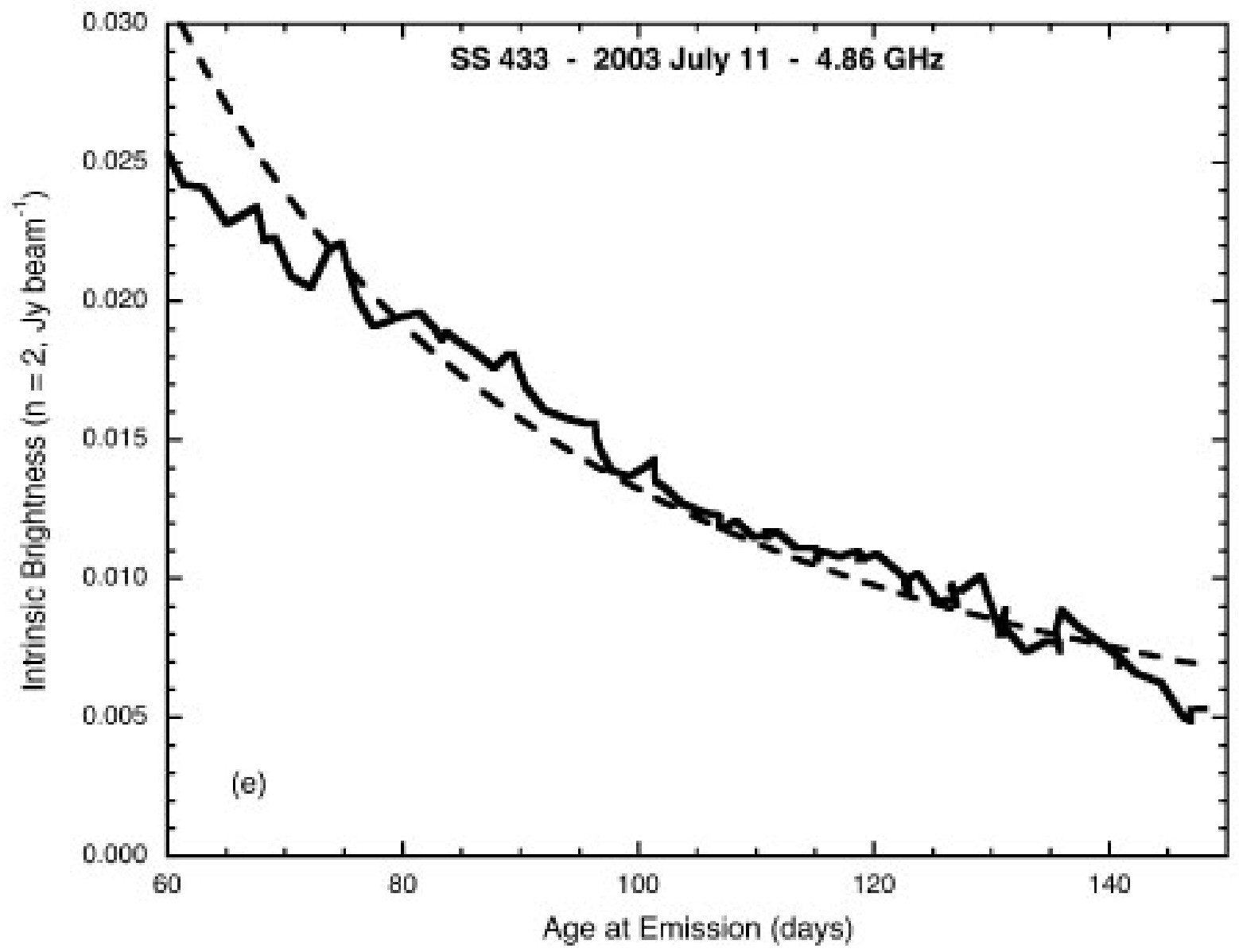}{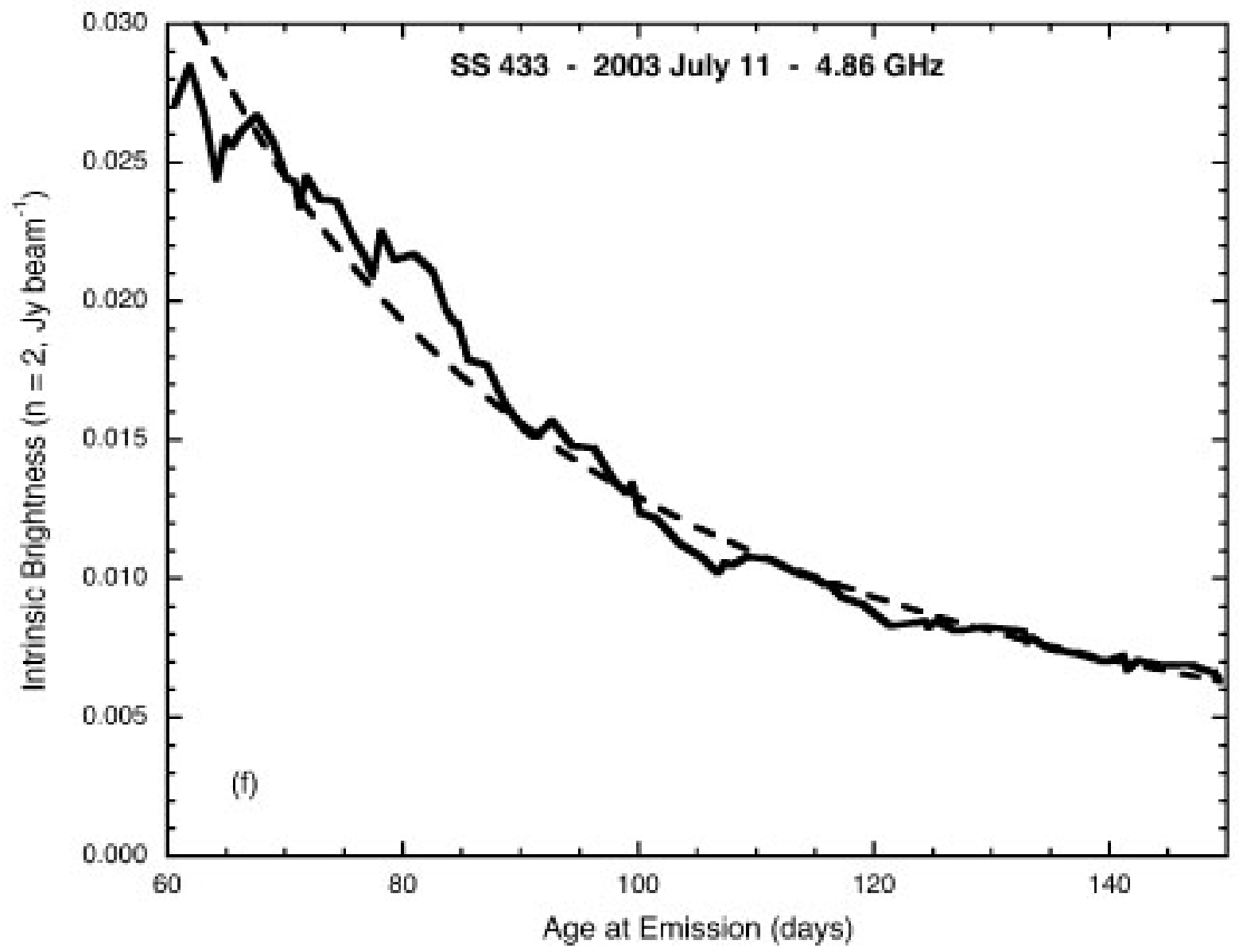}
\caption[]{Linear, exponential, and power law fits to the intrinsic brightness profiles of the jets of SS\,433 as functions of the age of the jet material (for $n = 2$), for ages in the range $60 \leq \tau \leq 150$~days (data from Figure~\ref{fig:CUN}). (a \& b) Linear fits; the east and west jets have linear half-lives of about 46 and 40~days, respectively. (c \& d) Exponential fits; the east and west jets have exponential half-lives of about  41 and 39~days, respectively. (e \& f) Power law fits; the east and west jets have power law indies of about $-1.7$ and $-1.8$, respectively. In all parts the fits are the broken lines.  As an example of the quality of the fits, the root-mean-square deviations from an exponential in the east and west jets are 0.8 and 1.0~mJy/beam, respectively.
\label{fig:Norm12LinearFits}}
\end{figure}

Assuming that the jets are dominated by aging rather than by variability in ejection properties, we can parameterize the physical processes that govern these different time ranges by fitting the normalized intensity curves with linear, exponential, and power law models. Figure~\ref{fig:Norm12LinearFits} shows fits to the intrinsic brightness $B$ of the jets for $n=2$ over the range $60 \leq \tau \leq 150$ days.  Exponential fits of the form $ B \propto e^{-\tau /T} $ yield half-lives of $T_{1/2} = T \ln{2} = 41$~days and 39~days for the east and west jets, respectively. Linear fits, while unrealistic for an entire jet, give half-lives of $T = -\bar{B}/2\dot{B}$ of 46~days and 40~days, and power law fits yield exponents of $-1.7$ and $-1.8$, for the east and west jets, respectively. Reassuringly, these timescales are very similar to the results of multi-epoch imaging that follows the aging of specific pieces of the jet (Paper~III, Paper~IV). The root-mean-square deviations of the exponential fits of $B(\tau)$ for the east jet and west jet are 1.0~mJy/beam and 0.8~mJy/beam, which suggests that a reasonable lower limit to the uncertainties in the total intensity curves is about 1~mJy/beam, comparable to the difference between using model curves with and without nutation and velocity variations (\S\ref{s:obs}). Given the limited span of these data, we are unable to distinguish definitively among these models; the exponentials appear superior, but this is not statistically significant.

For constant speed, as in SS\,433, models that predict intensity behavior with distance should be compared to data fit in age.  The model of a freely-expanding spherical cloud of magnetized relativistic plasma \citep{VanDerLaan}, which might be appropriate if the pieces of the jet behave as discrete  non-interacting components, predicts that in the optically-thin regime the resulting synchrotron total intensity falls off as $r^{-2p}$, where $p$ is the electron energy distribution exponent. Since here $p \simeq 2.4$,  this predicts a decay with a power law in component radius of index $\sim -4.8$. Were the components freely expanding we would have $r \propto \tau$, which would then be ruled out by the data.  In the conical jet model of \citet{HJ88}, the intensity of the jet falls off as $dI \propto z^{-m} \, dz$, where $z$ is distance down the jet, $m = (7p-1)/(6 + 6\delta) $, and $\delta = 0$ and $\delta = 1$ correspond to freely-expanding and slowed expansion cases, respectively. For $p = 2.4$, $\delta = 0$ corresponds to $ m = 2.6$, and $\delta = 1$ to $m = 1.3$. Power fits to our data in this age range lie between the two cases. We also fit exponentials and  power laws to the data for $250 \leq \tau \leq 800\mbox{ days}$, and find for both jets $T_{1/2} \simeq 80\mbox{ days}$ or $a \leq 4$ where $B \propto \tau^{-a}$, marginally incompatible with the freely expanding sphere model.

\section{CONCLUSIONS}
\label{s:conclude}

The principal results in this paper are:

\begin{enumerate}

\item We have used a deep VLA A-array image of SS\,433 at 4.86~GHz to study the intrinsic brightness profiles of the twin jets. 

\item Radiation from both jets is detected out to at least $6\arcsec$ from the core, corresponding to jet ages of about 800~days.

\item The observed brightnesses of the jets are strongly affected by projection effects and Doppler boosting.

\item Intrinsically the two jets are remarkably similar, and they are best described by Doppler boosting of the form $D^{2+\alpha}$, as expected for a continuous jet.

 \item  The intrinsic brightness of the jets behaves in a complex way that is not well described by single linear, exponential, or power law decay. 

\item During their first $\sim$150~days, the jet decays are well represented by linear or exponential functions of age, with  linear half-lives or exponential half-lives of about 40~days, the same for the two jets. Power law fits to the data in this age range give exponents of about $-1.8$. 
 
\item There is a transition region, corresponding to jet ages between about 150 and 250~days, during which the jets maintain roughly constant intrinsic brightnesses. This represents nearly one complete precession period. This also corresponds to about $ 150 < t < 250 \mbox{ days}$ in either jet. 

\item At later times the jet decay can be roughly fit as exponential functions of age, with exponential half-lives of about 80~days, or as power laws with indices of $a \leq 4$. 
   
\end{enumerate}

\section{Acknowledgments}

Part of this work is based on the undergraduate thesis of M.R.M. This material is based upon work supported by the National Science Foundation under Grants Nos.~0307531 and 0607453, and prior grants. Any opinions, findings, and conclusions or recommendations expressed in this material are those of the authors and do not necessarily reflect  the views of the National Science Foundation. D.H.R.\  gratefully acknowledges the support of the William R.\ Kenan, Jr.\ Charitable Trust. The National Radio Astronomy Observatory is a facility of the National Science Foundation, operated under cooperative agreement by Associated Universities, Inc. D.H.R.\ thanks Dale Frail and NRAO for their hospitality, and Vivek Dhawan, Amy Mioduszewski, and Michael Rupen for interesting conversations. Herman Marshall kindly provided us with the ephemeris for velocity variations and nutation, and very helpful comments. Rebecca Andridge contributed invaluable statistical expertise. We have made use of the CFITSIO and MFITSIO packages made available by the NASA Goddard SFC and Damian Eads of LANL, respectively. Finally, we thank the referee for a number of comments and questions that led to significant improvements and clarifications. 

Facilities: \facility{VLA (A array, data archive (experiment code AF403))}

 \section{APPENDIX 1}
 \label{s:app1} 

We used the  geometric model of \cite{HJ81b}, the precession parameters of \citet{Eik}, and the distance of \citet{LBG07}, augmented with jet velocity variations following \citet{BB05} and nutation following \citet{Katz82}, to determine the locus and kinematic properties of the jets and their appearance on the sky. The ephemeris used is listed in Table~\ref{tab:ephem}. 

Our convention for labeling a piece of jet that we see at a certain spot on the sky is as follows: We define its {\em birth epoch} $t$ as the time elapsed in the frame of the core since the material was born (ejected). Thus $t$ serves to label specific pieces of the jets. We also define its {\em age at emission} $\tau$ as the age of the same piece of material at the moment it emitted the photons that we detect at the same time as photons from the rest of the source.  Due to the finite speed of light, material that is ``behind'' the core and moving away from us had to emit its photons at a younger age in order that they arrive at the observer at the same time as photons from the core, and the reverse for matter ``in front.'' This means that oncoming material has $\tau > t$, receding material the opposite. In this convention, the core has $t = \tau = 0$, material with larger $t$ was born earlier than material with smaller $t$, material with larger $\tau$ emitted its photons at an older age than did material with smaller $\tau$, and material moving in the plane of the sky has $t = \tau$. Quantitatively, birth epoch $t$ and age at emission $\tau$ are related by
$$
\tau = \frac{t}{1-v_x/c},
$$
where $v_x$ is the component of material's velocity that is directed toward the observer. In SS\,433, each jet has pieces with both positive and negative $v_x$, as illustrated in Figures~\ref{fig:CUN} and \ref{fig:CNA}. Figure~\ref{fig:LTT} illustrates the light travel time effects in SS\,433. It shows the age at time of emission of photons $\tau$ for each part of the east and west jets, as a function of the birth epoch $t$ of each part of the jets. In the parts of the jets that are sufficiently bright that we can see them, the time differences $(t-\tau)$ range up to $\pm 200$~days, grow with distance from the core, produce the well-known distortion of the appearance of the jets, and affect our analysis. Note that at most places along the jets, the differences between $t$ and $\tau$ for the two jets are not equal for a given value of $t$.

\section{APPENDIX 2}
\label{s:app2} 

Here we address the question of why the jets behave such that their observed brightnesses vary as $D^{2+\alpha}$ rather than $D^{3+\alpha}$, as might be expected for isolated individual components such as those seen in VLBI imaging.\footnote{The appearance of isolated components in VLBI images could simply be due limited dynamic range.} There are four issues that should not be confused. (1) Are the jets a series of isolated components or a continuous fluid flow? (2) For these two cases, how does the observed total intensity depend on Doppler boosting in a jet whose locus is a helix but whose motion is radial? (3) Under what conditions can observations distinguish the the cases $n=2$ and $n=3$? (4) What do the data say about the SS\,433 jets?

A moving optically-thin synchrotron source is Doppler boosted and K-corrected by a factor of $D^{3+\alpha}$ because $I_\nu / \nu^3$ is a Lorentz invariant and $I_\nu \propto \nu^{-\alpha}$  \citep{RL} . 
In the case of a continuous jet we want to know the flux density of a segment of the jet defined by the observing beam. As shown by \cite{LindBlandford, Sikora,DeYoung}, and others, not all the material in that segment of jet at a given instant contributes to a particular image. This is purely a light travel time effect. The fraction of  the jet segment that contributes is $(1 - \beta\cos\theta) = 1/\gamma D$. The effect of this is that the boost factor is modified from $D^{3+\alpha}$ to $D^{2+\alpha}/\gamma$; in other words, $I_\nu^{jet} = \gamma I_\nu^{obs}/ D^{2+\alpha}$. This calculation assumes a straight jet with the velocity along the jet
direction.

In SS\,433 the locus of the jet is a helix even though the motion of the jet material is radial. 
This means that the time delay between the arrival of photons from the near and far ends  of the jet is no longer $\Delta t = R \cos(\theta)/c$, where $R$ is the length of the part of the jet defined by the beam and $\theta$ is the angle between the jet velocity and the line of sight. Instead, it is $\Delta t \simeq R \cos(\eta)/c$, where $\eta$ is the angle between the tangent to the jet locus and the line of sight. In addition, the rate at which material is added to the observed part of the jet is no longer proportional to $v_{jet}$, but is instead determined by the component of the jet velocity in the direction of the locus of the helix. We have made such calculations using the known kinematics of each part of the jet, and the results are very similar to the continuous straight jet case, that is, with effective boosts of approximately $D^{2+\alpha}$, as observed. Critical to this analysis is the fact  that the Lorentz factors of each part of the jets are the same.

If the jets are each a series of components, then whether the total intensity is boosted by $D^{2+\alpha}$ or $D^{3+\alpha}$ depends on the spacing of the components relative to the spatial scale of the resolution of the imaging. If the components are sufficiently close, then there are always components that do not appear in the beam but belong there, and components that appear there but are not, just as for a continuous fluid flow, and $D^{2+\alpha}$ is appropriate. If the spacing is so great that during the light travel time from the back of the jet to the front no component moves into or out of the beam, then $D^{3+\alpha}$ is appropriate. However, depending on the properties of the components and the angular resolution of the observation, any value between $n=2$ and $n=3$ can be appropriate. For the current observation, the data strongly suggest that $n=2$ for SS\,433, indicating that the jets are either continuous, or if composed of discrete components, then there are many in the beam at any time.

\begin{deluxetable}{lllc}

\tablecaption{Ephemeris for SS\,433.\label{tab:ephem}}
\tablewidth{0pt}
\tablecolumns{4}
\tablehead{\colhead{Property} & \colhead{Symbol\tablenotemark{a}} & \colhead{Value}  }

\startdata

Speed & $\beta = v/c$ & $0.2647$ \\
Precession Period & $P$ & 162.375~d  \\
Reference Epoch (JD) & $t_{ref}$ & 2443563.23  \\
Precession Cone Opening Angle & $\psi$ & $20.92^{\circ}$  \\
Precession Axis Inclination & $i$ & $78.05^{\circ}$ \\
Precession Axis Position Angle & $\chi + \pi/2$ & $98.2^{\circ}$  \\
Sense of Precession & $s_{rot}$ & $-1$  \\
Distance & $d$ & $5.5~\mbox{kpc}$  \\ 
Orbital Period Reference Epoch (JD) & \ldots & 2450023.62 \\
Orbital Period & \ldots & 13.08211~d \\
Nutation Reference Epoch (JD) & \ldots & 2450000.94 \\
Nutation Period & \ldots & 6.2877~d \\
Nutation Amplitude & \ldots & 0.009	\\
$\beta_{orb}$ & \dots & 0.0066	\\
$\beta_{orb \, phase}$ & \ldots & 4.7 \\
\enddata
\tablenotetext{a}{As defined in \cite{HJ81b}.}
\end{deluxetable}

\clearpage


\epsscale{.70}
\epsscale{1.0}

\clearpage


\clearpage

\clearpage

\clearpage

\clearpage

\clearpage

\clearpage

\clearpage

\clearpage

\clearpage

\epsscale{0.65}

\end{document}